\documentclass[english,a4paper]{article}
\usepackage{fullpage}
\usepackage{graphicx}
\usepackage{amssymb}
\usepackage{amsmath}
\usepackage{bibentry}
\usepackage{ulem}
\usepackage{xcolor}
\usepackage{authblk}

\makeindex

\begin{document}
\date{}
\title{Fully integrative data analysis of NMR metabolic fingerprints with comprehensive patient data: a case report based on the German Chronic Kidney Disease (GCKD) study}

\author[1,$\dagger$]{Helena U. Zacharias}
\affil[1]{Institute of Computational Biology, Helmholtz Zentrum M\"unchen, Ingolst\"adter Landstra\ss e 1, 85764 Neuherberg, Germany.}
\affil[$\dagger$]{authors contributed equally to this work}
\author[2,$\dagger$]{Michael Altenbuchinger}
\affil[2]{Statistical Bioinformatics, Institute of Functional Genomics, University of Regensburg, Am Biopark 9, 93053 Regensburg, Germany.}
\author[3]{Stefan Solbrig}
\affil[3]{Department of Physics, University of Regensburg, Universit\"atsstra\ss e 31, 93053 Regensburg, Germany.}
\author[3]{Andreas Sch\"afer}
\author[1]{Mustafa B\"uy\"uk\"ozkan}
\author[4,5]{Ulla T. Schulthei\ss}
\affil[4]{Institute of Genetic Epidemiology, Department of Biometry, Epidemiology, and Medical Bioinformatics, Faculty of Medicine and Medical Center, University of Freiburg, 79106 Freiburg, Germany.}
\affil[5]{Department of Nephrology, Medical Center, University of Freiburg, 79106 Freiburg, Germany.}
\author[4]{Fruzsina Kotsis}
\author[4]{Anna K\"ottgen}
\author[1,6]{Jan Krumsiek}
\affil[6]{Institute of Computational Biomedicine, Weill Cornell University, New York, NY 10021, USA.}
\author[1,7]{Fabian J. Theis}
\affil[7]{Department of Mathematics, Technische Universit\"at M\"unchen, Germany.}
\author[2]{Rainer Spang}
\author[8]{Peter J. Oefner}
\affil[8]{Institute for Functional Genomics, University of Regensburg, Am Biopark 9, 93053 Regensburg, Germany.}
\author[8]{Wolfram Gronwald}
\author[9]{GCKD study investigators}
\affil[9]{GCKD study investigators are listed in the acknowledgement.}

\maketitle

\begin{abstract} 
Omics data facilitate the gain of novel insights into the pathophysiology of diseases and, consequently, their diagnosis, treatment, and prevention. To that end, it is necessary to integrate omics data with other data types such as clinical, phenotypic, and demographic parameters of categorical or continuous nature. Here, we exemplify this data integration issue for a study on chronic kidney disease (CKD), where complex clinical and demographic parameters were assessed together with one-dimensional (1D) ${}^1$H NMR metabolic fingerprints. 

Routine analysis screens for associations of single metabolic features with clinical parameters, which requires confounding variables typically chosen by expert knowledge to be taken into account. This knowledge can be incomplete or unavailable.

The results of this article are manifold. We introduce a framework for data integration that intrinsically adjusts for confounding variables. We give its mathematical and algorithmic foundation, provide a state-of-the-art implementation, and give several sanity checks. In particular, we show that the discovered associations remain significant after variable adjustment based on expert knowledge. In contrast, we illustrate that the discovery of associations in routine analysis can be biased by incorrect or incomplete expert knowledge in univariate screening approaches. Finally, we exemplify how our data integration approach reveals important associations between CKD comorbidities and metabolites. Moreover, we evaluate the predictive performance of the estimated models on independent validation data and contrast the results with a naive screening approach.
\end{abstract}

\section{Introduction}

The advent of new omics technologies, offering high coverage at an affordable price, has changed the landscape of large-scale clinical studies. More and more population-based trials collect not only information about phenotypes, traditional clinical parameters, and demographic variables, but also extensive omics data, e.g. KORA \cite{holle2005kora, illig2010genome}, and TwinsUK \cite{moayyeri2012cohort}. This typically results in a large set of complex patient data, which needs to be statistically analyzed. 

One example of such a large-scale, population-based study is the German Chronic Kidney Disease (GCKD) study, comprising about 5200 chronic kidney disease (CKD) patients. CKD constitutes one of the largest burdens on the world's health system \cite{jha2013chronic}, with about 200 cases per million per year in many countries \cite{levey2012chronic}, and a global occurrence of 50\% in high-risk subpopulations \cite{eckardt2013evolving}. As it progresses, CKD leads to a large number of adverse clinical symptoms, and in some cases may progress to complete renal failure, called end-stage renal disease (ESRD) \cite{kuhlmann2008nephrologie}. It is linked to elevated all-cause and cardiovascular mortality, acute kidney injury (AKI), cognitive decline, anemia, mineral and bone disorders, and fractures \cite{jha2013chronic,kuhlmann2008nephrologie,kdw2013kidney}. Moreover, it is often accompanied by numerous other systemic diseases, such as cardiovascular disease, hypertension, obesity, and diabetes \cite{eckardt2013evolving,chawla2014acute,o2014kidney}.

The GCKD study was designed to improve our understanding of the causes, course, and risk factors of progressive kidney insufficiency \cite{eckardt2011german}. To that end, demographic, phenotypic, and clinical parameters of approx. 5200 CKD patients were assessed \cite{eckardt2011german, titze2014disease}. Additionally, NMR metabolic fingerprints of plasma specimens collected at the study baseline were acquired for almost the complete cohort. Metabolic markers are expected to better reflect the current state of the kidney than, e.g., genomic, transcriptomic, or proteomic markers \cite{wishart2006metabolomics}. In total, approximately 900 parameters of diverse data sources, which potentially have complex dependencies amongst each other, have been assembled. Here, we aim to identify, in the best case yet unknown, relationships between these complex layers of information. Particularly, we aim to uncover metabolic features related to CKD and its diverse comorbidities. 

Data integration tries to detect such relationships. However, it presents both a conceptual and a practical challenge.  Conceptually, we are faced with complex, often synergistic effects between variables within and across different layers of information, where the variables are potentially of different data types. Practically, we have to deal with a large-scale dataset covering close to thousand variables and several thousand measurements, leaving the researcher with a lot of possible hypotheses that require extensive validation. The practical and conceptual issues are inseparably connected. Assuming more complex and realistic models requires more parameters and their estimation becomes computationally more and more challenging. 

Numerous methods have been proposed to analyze complex datasets, ranging from metabolome-wide association studies and multivariate statistical analysis to data-driven network-based approaches \cite{krumsiek2016computational, zierer2015integration}. Metabolome-wide association studies are applied routinely to identify associations between metabolites and phenotypes, but they inherently ignore complex, multivariate relationships between variables. Approaches based on probabilistic graphical models can reveal complex relationships, but they are usually limited to one specific data type, e.g., Gaussian  Graphical  Models (GGMs) for continuous (Gaussian) variables or discrete Markov Random Fields (dMRF) for categorical variables \cite{lauritzen1996graphical}. More recently, probabilistic graphical models have been extended to include different data types simultaneously \cite{lauritzen1996graphical, lee2015learning}, although such methods have not yet entered biomedical research. 

Univariate screening approaches ignore effects of confounding variables. These can be either incorporated by correcting the data or by adapting the statistical test. Formally, this corresponds to estimating the partial correlation coefficient $\rho_{XY\cdot \textrm{Z}}$, which denotes the partial correlation between $X$ and $Y$ given the confounders $Z$. If $\rho_{XY\cdot \textrm{Z}}\ne0$, then there is an association between $X$ and $Y$ given $Z$, where the size and sign of $\rho_{XY\cdot \textrm{Z}}$ reflects the strength and sign of an association, respectively. An inherent problem in the estimation of $\rho_{XY\cdot \textrm{Z}}$ is that it is not \textit{a priori} clear which variables to include in $Z$. To solve this issue, we need to estimate $\rho_{XY\cdot \textrm{Z}}$, where $Z$ includes all assembled variables except $X$ and $Y$, ideally containing the complete confounder set. 

Here, we estimate $\rho_{XY\cdot \textrm{Z}}$, where $X$, $Y$, and $Z$ are the comprehensive patient data assembled within the GCKD study. The included variables are either continuous or categorical. Thus this analysis requires the estimation of a so-called Mixed Graphical Model (MGM) \cite{lauritzen1996graphical, lee2015learning}. We first show how this integrative analysis reveals known relationships between variables. Second, we illustrate that the discovery of associations in univariate screening approaches is biased by incorrect or incomplete expert knowledge. Third, we demonstrate that our data integration approach overcomes this issue. Finally we give an example, where important associations between CKD comorbidities and metabolites are revealed. Here, we show that the discovered associations remain significant after variable adjustment based on expert knowledge. Throughout the article, we substantiate our findings by evaluating the predictive performance of the estimated model on validation data.
 
\section{Materials and methods} 

\subsection{Mixed Graphical Models}
MGMs are undirected probabilistic graphical models, where the conditional dependencies between different variables, the so-called nodes or vertices, are represented as edges in a network. Thus, two nodes are connected by an edge only if their association or interaction cannot be explained by any other node in the graphical model, or equivalently, if their partial correlation coefficient is unequal zero. Consequently, probabilistic graphical models eliminate spurious associations between variables and can potentially reveal new associations adjusted for all other variables in the dataset.

\subsection{Algorithmic implementation}
For the current dataset, which included $879$ variables and $3705$ patients (measurements), we needed to estimate, in total, $388521$ edge weights. 

Therefore, we implemented an efficient algorithm to estimate MGMs as described in the supplementary methods section \ref{MGMs}. Here, we used the pseudo-log-likelihood method of \cite{lee2015learning}, made use of proximal algorithms to include a LASSO penalty on the parameters \cite{parikh2014proximal}, and we performed the proximal gradient descent using Nesterov acceleration and adaptive restarts \cite{nesterov2007gradient, o2015adaptive}. The respective Python scripts to estimate MGMs are available upon request.

\subsection{Data integration workflow}
\begin{figure}[t!]
\centering
\includegraphics[width=.8\textwidth]{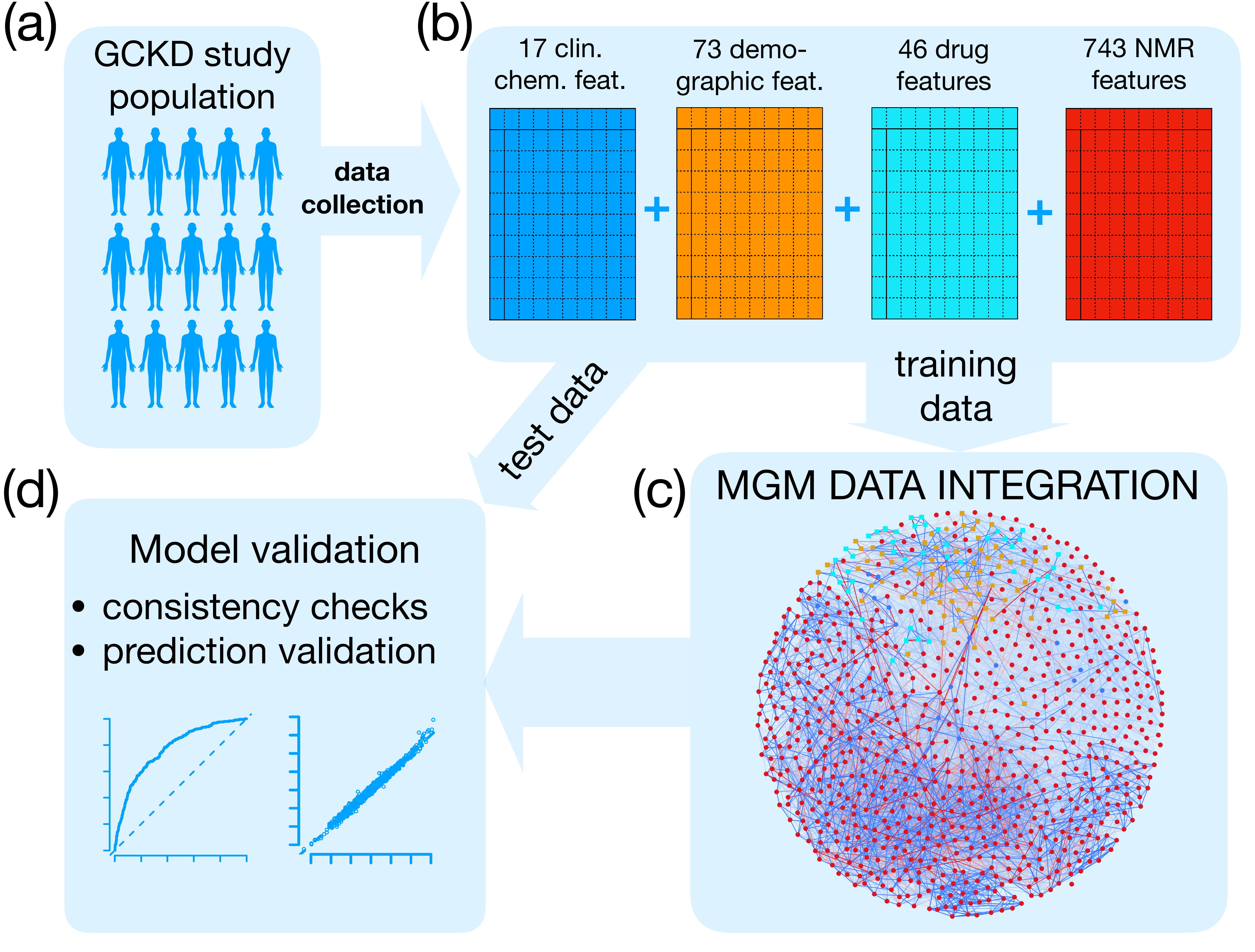}
\caption{\label{workflow}\footnotesize{Scheme of the  Mixed  Graphical Model  (MGM)  data  integration approach. (a) Of the data ascertained from the GCKD  study population, (b) a  total of 17 clinical chemistry parameters (blue),  73 demographic parameters (orange),  46 drug  treatment parameters (cyan),  and  743 NMR spectral features (red) were chosen.  The complete  dataset  was split  into  a training and a test cohort, respectively. The first (c) was used  to estimate an MGM, modeling all conditional dependencies between  all variables, whereas  the latter (d) was used for MGM model validation. In the network representation of the estimated MGM, blue nodes represent  clinical  chemistry parameters, orange  nodes  represent  demographic variables, cyan  nodes  represent drug  treatment information, and  red  nodes  correspond to  NMR  buckets.  Continuous variables are represented as circles  and  discrete variables as  rectangles. Positive and negative associations are  shown  as  blue  and  red  edges, respectively. The strength of the  association, i.e., the weight of the corresponding coefficient, is encoded  by the  edge width. 
}}
\end{figure}

A schematic illustration of our data integration workflow is shown in Figure \ref{workflow}. Starting point is the GCKD study, Figure \ref{workflow}a. 
Here, we included a total of 3705 GCKD study participants. The remaining study participants had to be excluded due to missing demographic, clinical, and/or metabolic data. We generated a training set of 2470 study participants, corresponding to 2/3 of the complete cohort, and a test set of 1235 study participants, respectively, compare to Figure \ref{workflow}b. For the current study, we included 17 clinical chemistry parameters, 73 demographic parameters, and 46 different drug treatments, all assessed at the study baseline. Each of the three different information layers is represented by a blue, orange, and cyan schematic data matrix in Figure \ref{workflow}b, respectively. These parameters, their distributions, assessment, as well as transformation information are given in Supplementary Table S1. For each study participant, a one-dimensional (1D) ${}^1$H nuclear magnetic resonance (NMR) spectrum of a baseline EDTA plasma specimen was acquired. In total, 743 metabolic features were extracted from the spectra, as described in the supplementary methods section \ref{dataPrepro}. This information layer is represented by the red data matrix in Figure \ref{workflow}b. In summary, the complete variable space included 879 variables, with 768 continuous and 111 discrete variables, respectively.

Next, the MGM was estimated on the training data, Figure \ref{workflow}c, and validated on independent test data, Figure \ref{workflow}d. Here and throughout the article, clinical chemistry parameters are represented as blue nodes, demographic variables as orange nodes, drug treatment information as cyan nodes, and NMR buckets as red nodes. Positive and negative associations are shown as blue and red edges, respectively, where the edge width encodes the absolute edge strength. Continuous variables are shown as circles and discrete variables as rectangles.

\section{Results}
\subsection{MGM data integration reveals known associations and combines them to reliable prediction models}
First, we will present several sanity checks for our MGM approach. We will illustrate how the MGM reveals known associations in the context of the two renal function markers urinary albumin-to-creatinine ratio and the glomerular filtration rate, for two frequent comorbidities, i.e. diabetes and gout, and for a lifestyle factor, i.e. alcohol consumption. 

\begin{figure}[t!]
\centering
\includegraphics[width=.4\textwidth]{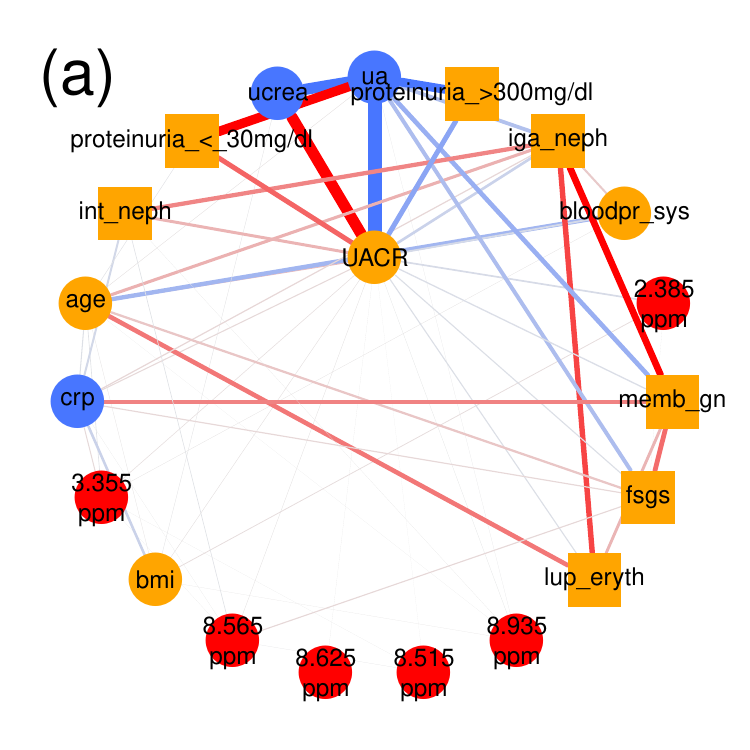}
\includegraphics[width=.4\textwidth]{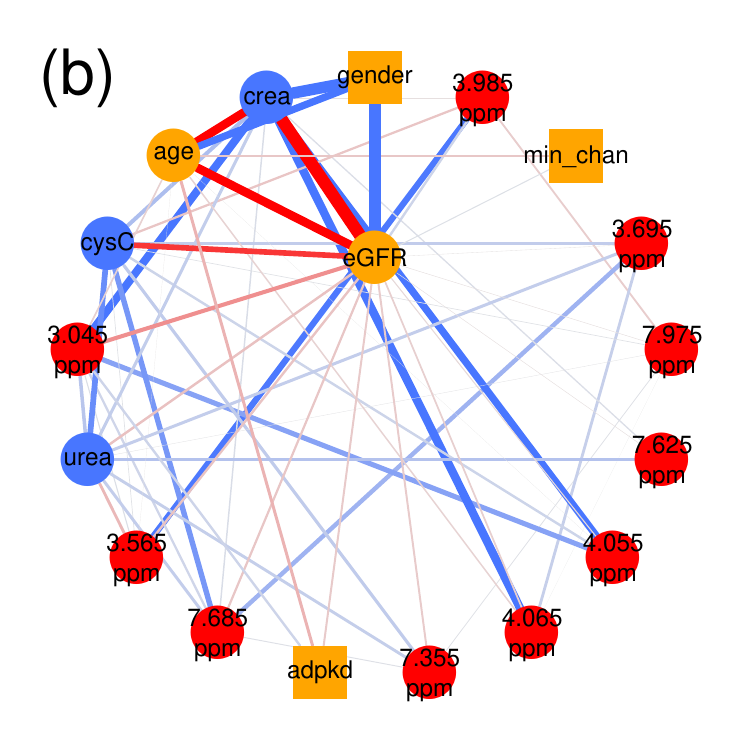}
\caption{\label{neiuacr}{\footnotesize{(a) First order neighborhood of urinary albumin-to-creatinine-ratio (\textit{UACR}). The first order neighborhood of a node comprises all other nodes which have a direct association/edge with this particular node. Positive associations are represented as blue, negative associations as red edges, respectively. The strength of the estimated association is encoded by the edge width. The edges are ordered according to their strength in a clock-wise manner for positive, and in an anti-clock-wise manner for negative associations, respectively. \textit{UACR} is positively associated with urinary albumin (\textit{ua}) (edge weight = 22.85) and negatively associated with urinary creatinine (\textit{ucrea}) (edge weight = -7.19), respectively. The positive edge between \textit{ua} and \textit{ucrea} (edge weight = 7.31) can be explained within the conditional dependence framework (see text). Two levels of proteinuria are connected to both \textit{UACR} and \textit{ua}, respectively. High level of protein in the patient's urine (\textit{proteinuria $>$ 300mg/dL}) is positively correlated, whereas proteinuria of less than 30mg/dL (\textit{proteinuria $<$ 30mg/dL}) is negatively correlated with \textit{UACR} (edge weight = 0.573 and edge weight = -0.564, respectively) and \textit{ua} (edge weight = 2.873 and edge weight = -3.53, respectively), respectively. Moderate proteinuria (\textit{30mg/dL $<$ proteinuria $<$ 300mg/dL}) was considered as the baseline. (b) First order neighborhood of eGFR values estimated by the CKD-EPI equation based on serum creatinine ($eGFR$). $eGFR$ is strongly negatively associated with serum creatinine ($crea$) (edge weight = -11.185), strongly positively associated with male gender ($gender$) (edge weight = 7.514), and negatively associated with age ($age$) (edge weight = -2.713). Negative associations are revealed between $eGFR$ and serum cystatin C values ($CysC$) (edge weight = -0.764), and the NMR bucket at 3.045ppm (edge weight = -0.374), corresponding to creatinine, respectively. A complete list of all abbreviations for the clinical parameters can be found in Supplementary Table S1.}}}
\end{figure}

\paragraph{Urinary albumin-to-creatinine ratio}
The urinary albumin-to-creatinine ratio (\textit{UACR}) and the glomerular filtration rate (GFR) are the two most important markers of renal function routinely assessed in CKD patients. Figure \ref{neiuacr}a visualizes the first order neighborhood of \textit{UACR}. The first order neighborhood of a node comprises all other nodes that have a direct association/edge with this particular node. The MGM trained on the training set connects \textit{UACR} with urinary albumin (\textit{ua}) by a strong positive edge, and with urinary creatinine (\textit{ucrea}) by a strong negative edge, respectively. This correctly reflects the definition of \textit{UACR}, $\log_2(\textit{UACR})=\log_2(\textit{ua}/\textit{ucrea})=\log_2(\textit{ua})-\log_2(\textit{ucrea})$.

A third positive edge appears between \textit{ua} and \textit{ucrea}. This edge can be explained within the conditional dependence framework: two variables are conditionally dependent on each other, if both variables simultaneously change upon holding all other variables fixed. Let the value of \textit{UACR} be fixed and change the value of \textit{ua}. In order to fulfill the requirement that the \textit{UACR} stays fixed, urinary creatinine has to change accordingly to compensate the changes of urinary albumin and, thus, a positive edge appears between \textit{ua} and \textit{ucrea}. 

We further observe positive edges between severe proteinuria (\textit{proteinuria $>$ 300mg/dL}) and \textit{ua} as well as \textit{UACR}, respectively, correctly stating that patients with high urinary levels of albumin and accordingly high \textit{UACR} values are diagnosed with proteinuria. Correspondingly, absent or mild proteinuria (\textit{proteinuria $<$ 30mg/dL}) is negatively associated with urinary albumin and \textit{UACR}. Here, intermediate proteinuria (\textit{30mg/dL $<$ proteinuria $<$ 300mg/dL}) was considered as the baseline.

\begin{figure}[t!]
\centering
\includegraphics[width=.24\textwidth]{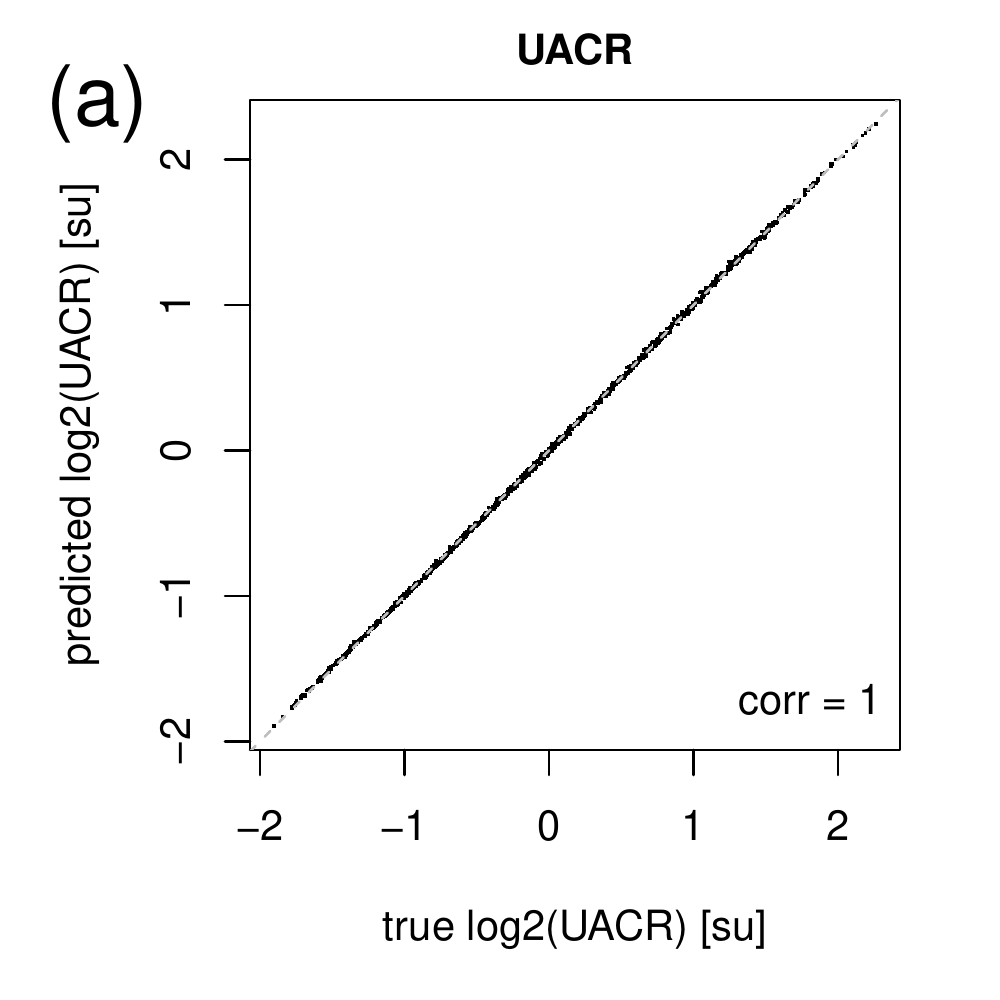}
\includegraphics[width=.24\textwidth]{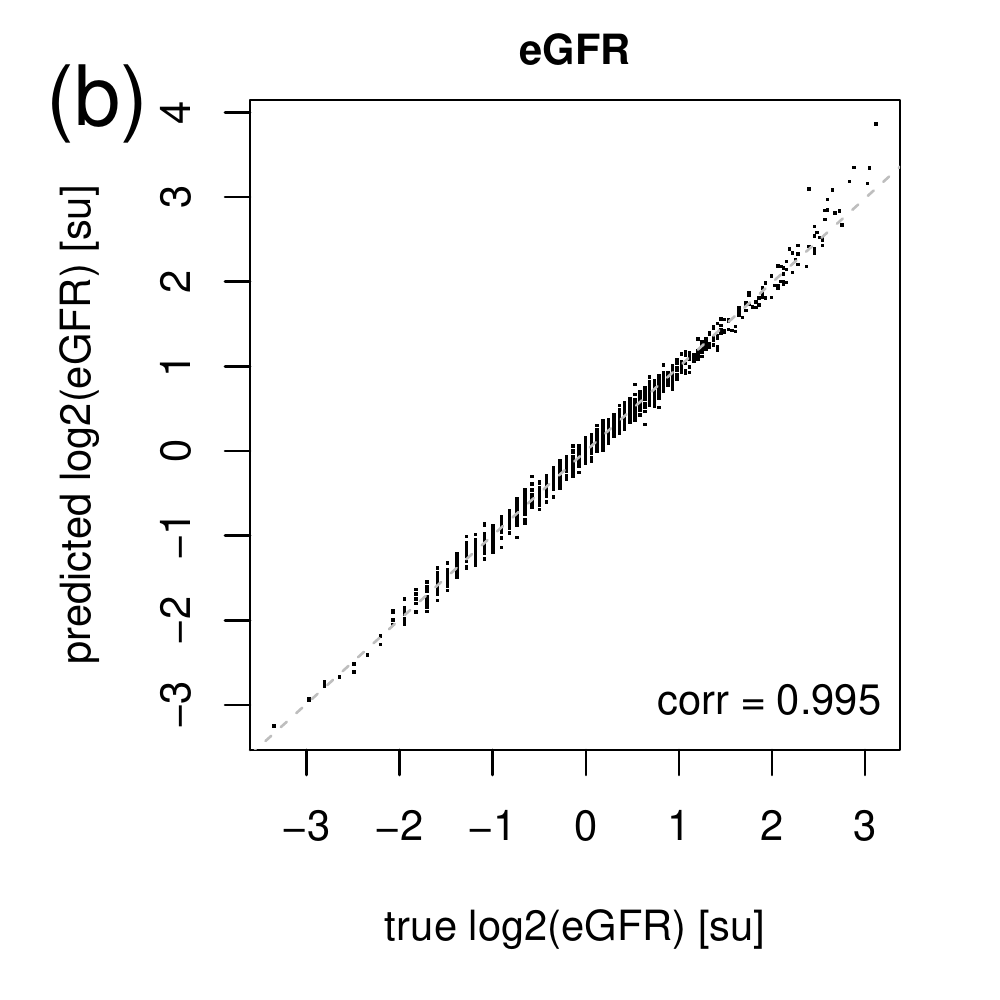}\\
\includegraphics[width=.24\textwidth]{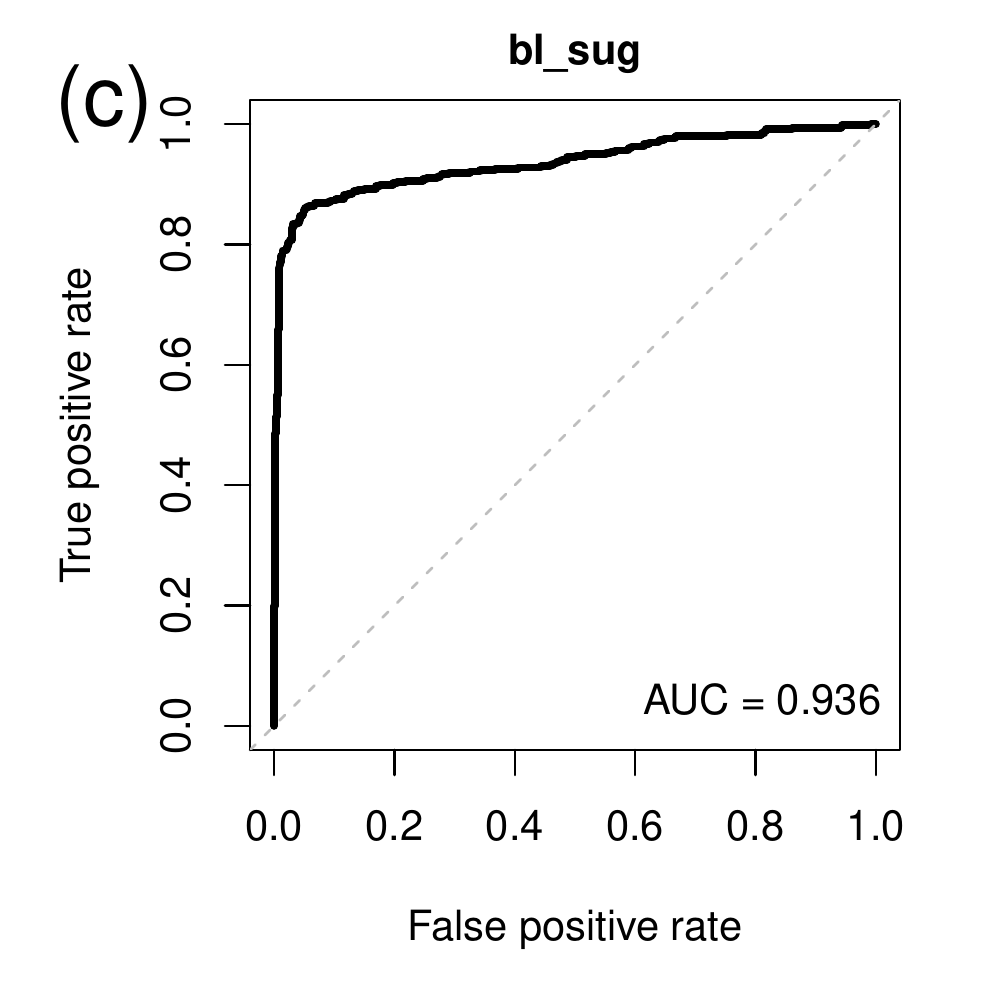}
\includegraphics[width=.24\textwidth]{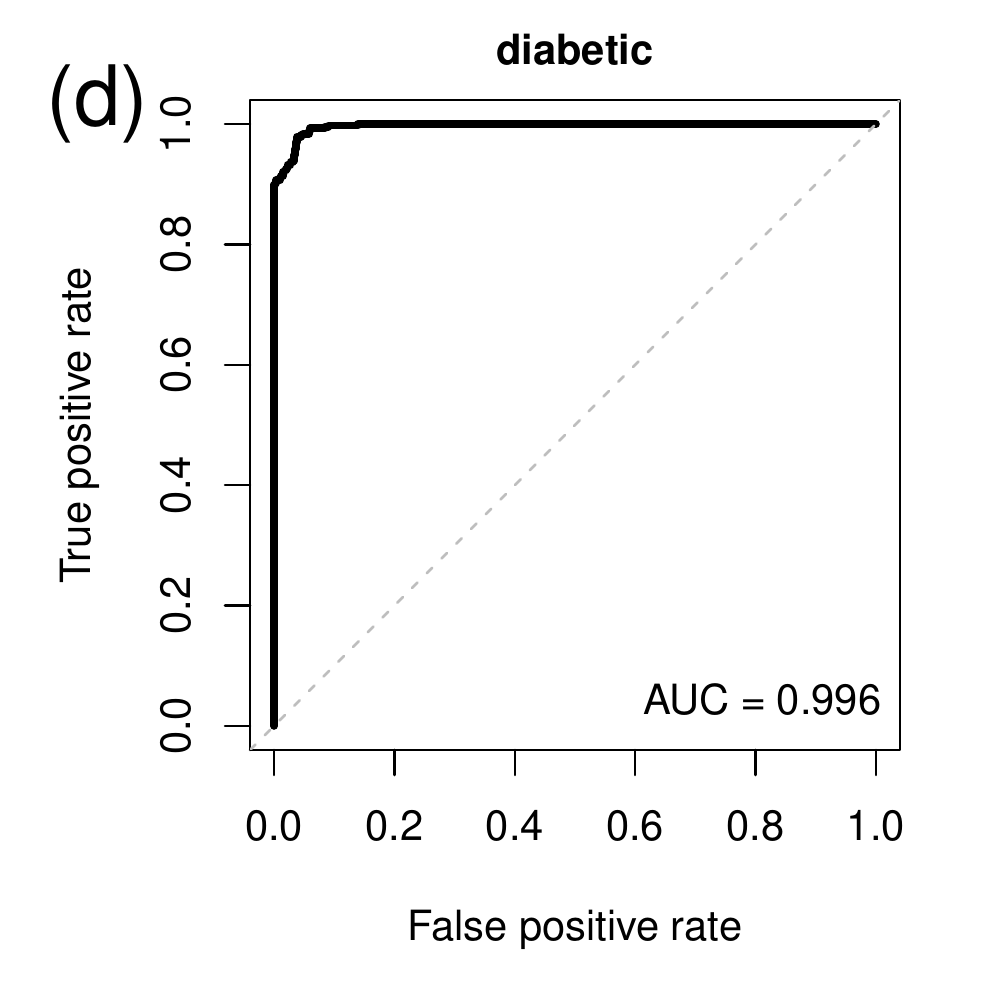}
\caption{\label{pred_uacr_egfr}{\footnotesize{(a) The diagram shows the predictions of \textit{UACR} on the $y$-axis (in standard units [su]) based on the neighbors of \textit{UACR}, compare to Figure \ref{neiuacr}a, on independent test data compared to the true values plotted on the $x$-axis. (b) Shows the analogous plot for \textit{eGFR}. In both scenarios, predictions agree almost perfectly with the true values as indicated by the correlation coefficients $cor$ between true and predicted values given in the lower right corners. The receiver operating characteristic (ROC) curve for predicting elevated blood sugar based on its neighborhood, as shown in Figure \ref{neiblutzu}a, is shown in (c). The $x$-axis here represents the false, whereas the $y$-axis represents the true positive rate, respectively. The dashed line gives the diagonal, corresponding to the predictive performance of a randomly generated model. In the lower right corner, the area under the ROC curve (AUC), an indicator of the predictive power of a classifier, is given. A perfect classifier would achieve an AUC of 1 on independent test data, whereas a randomly generated classifier with no predictive power would achieve an AUC of 0.5, respectively. (d) shows the ROC curve for the neighborhood model of the medical diagnosis of a patient as being diabetic.}}}
\end{figure}

In a second step, we validated the predictive performance of the detected neighborhood. For this purpose, we used the respective edge weights as a linear signature that predicts \textit{UACR}. As shown in Figure \ref{pred_uacr_egfr}a, the identified neighborhood is almost perfectly predictive of \textit{UACR} on independent test data with a correlation coefficient between true and predicted \textit{UACR} values of $cor\sim1$.

\paragraph{Glomerular filtration rate (GFR)}
The GFR defines the fluid volume filtered by the glomeruli per unit time. Several formulas are available to estimate the GFR. Here, we included estimated GFR (\textit{eGFR}) values calculated using the CKD-EPI equation \cite{levey2009new}. 

CKD-EPI takes as input serum creatinine ($crea$), age, gender, and race. In Figure \ref{neiuacr}b, we show the first order neighborhood of \textit{eGFR}, which correctly identified age, gender, and serum creatinine. Race was not included as a variable, as all GCKD study participants were Caucasian, and, thus, the respective connection could not be observed.

In addition, we observed associations of eGFR with cystatin C (\textit{cysC}), and NMR buckets assigned to creatinine ($3.045$ ppm, $4.055$ ppm, and $4.065$ ppm). Serum cystatin C, besides serum creatinine, is used as an endogenous marker to estimate GFR, e.g. in the CKD-EPI equation based on cystatin C \cite{stevens2008estimating}, which explains the found association between cystatin C and eGFR.

In Figure \ref{pred_uacr_egfr}b, we show how the neighborhood of \textit{eGFR} predicts \textit{eGFR} on test data. We observe that values agree with a correlation coefficient of $cor=0.995$.

\paragraph{Diabetes}

One of the most common comorbidities of CKD is type-2 diabetes mellitus (T2DM). Untreated T2DM is characterized by high blood glucose. In this study, blood sugar is included as the discrete variable ``elevated blood sugar yes/no'' (\textit{bl\_sug}).

\begin{figure}[t!]
\centering
\includegraphics[width=0.49\textwidth]{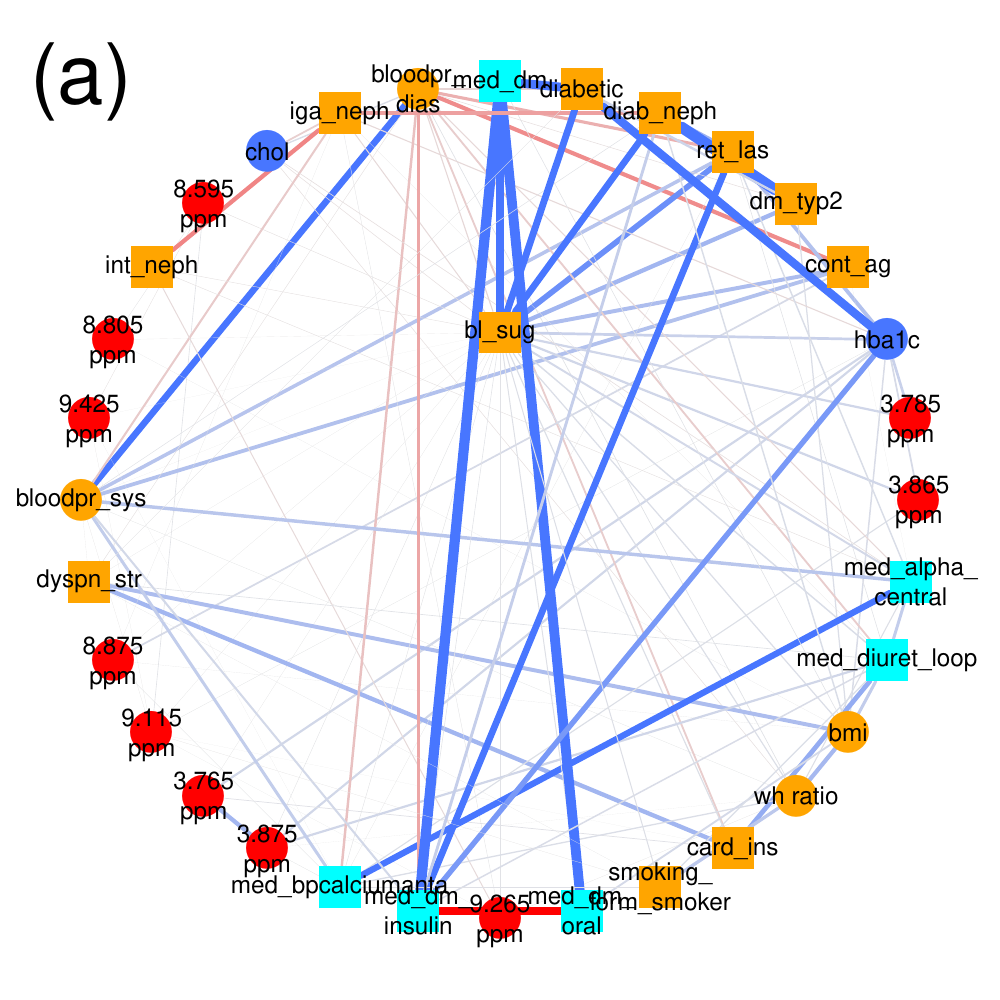}
\includegraphics[width=0.49\textwidth]{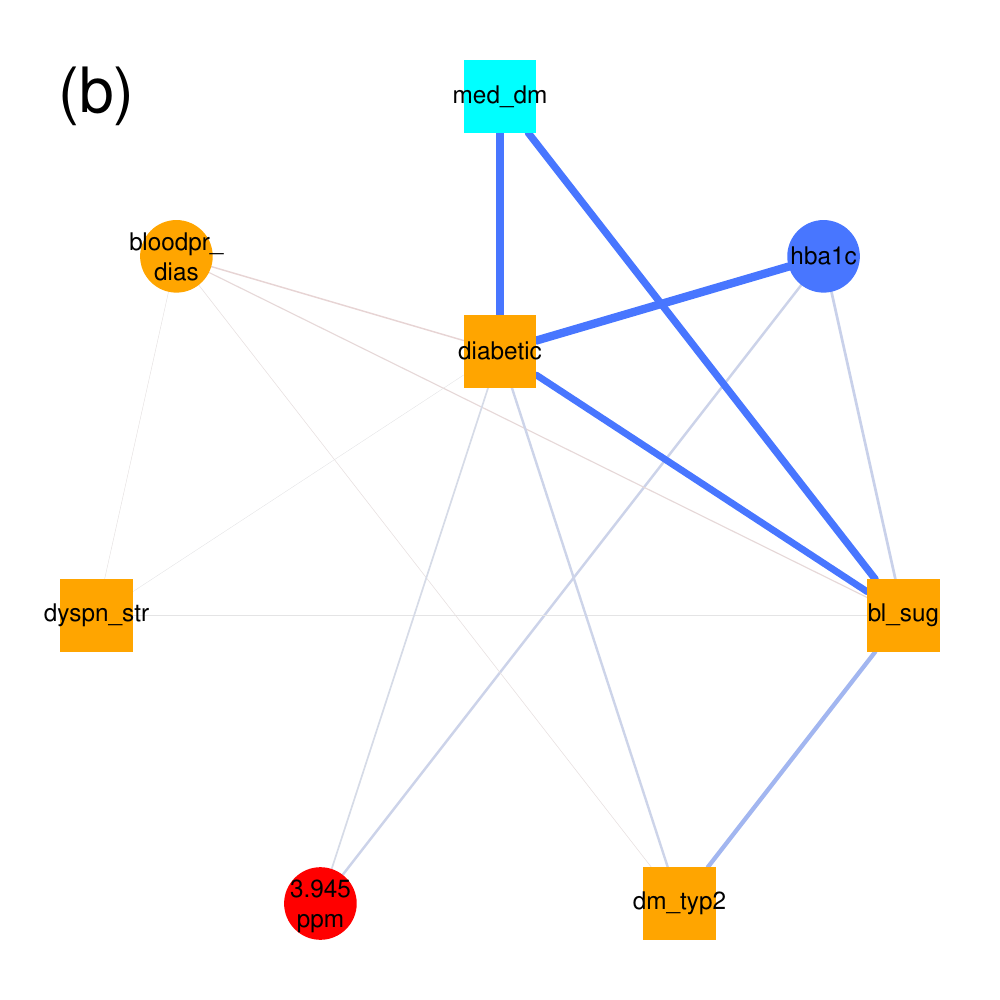}
\caption{\label{neiblutzu} {\footnotesize{(a) First order neighborhood of elevated blood sugar (\textit{bl\_sug}) and (b) classification as diabetic patient (\textit{diabetic}). Strong associations can be observed between \textit{bl\_sug} and diabetes medications ($med\_dm$) (edge weight = 1.52), $diabetic$ (edge weight = 1.27), and diabetic nephropathy ($diab\_neph$) (edge weight = 1.15), respectively. Other strong associations are present between $diabetic$ and \textit{med\_dm} (edge weight = 2.89), and the HbA1c value ($hba1c$) (edge weight = 2.366), respectively, as well as between 2 NMR buckets at 3.785ppm (edge weight = 0.136) and 3.865ppm (edge weight = 0.134), both corresponding to D-glucose, and \textit{bl\_sug}, and between $diab\_neph$ and classification as type-2 DM patient (\textit{dm\_typ2}) (edge weight = 7.1) and retinal laser therapy due to diabetes ($ret\_las$) (edge weight = 0.47), respectively.}}}
\end{figure}

The MGM correctly identifies NMR buckets corresponding to D-glucose signals, namely 3.785 ppm, 3.865 ppm, 3.875 ppm, and 3.765 ppm, in the first-order neighborhood of ``elevated blood sugar yes/no'', Figure \ref{neiblutzu}a. The three strongest connections are between elevated blood sugar and diabetes medications (\textit{med\_dm}), the classification as diabetes patient (\textit{diabetic}), and diabetic nephropathy (\textit{diab\_neph}), respectively.

In Figure \ref{neiblutzu}b, we show the first-order neighborhood of $diabetic$, which exhibits a strong connection to diabetes medications (\textit{med\_dm}) and glycated hemoglobin (HbA1c) (\textit{hba1c}). This reflects the definition of a diabetic patient within this study: a patient is defined as diabetic, if he had an HbA1c value $\ge$ 6.5\% or took at least one medication with an active component classified as ``A10'', i.e. insulin and other diabetes medications.  HbA1c reflects the two to three month average plasma glucose level \cite{sherwani2016significance} and is here associated with elevated blood sugar independently of diagnosed diabetes. 

Other strong connections were observed between blood sugar and retinal laser therapy due to diabetes (\textit{ret\_las}), as well as classification as T2DM patient (\textit{dm\_typ2}). All three variables are positively connected to diabetic nephropathy (\textit{diab\_neph}), as a study participant is stated to suffer from diabetic nephropathy, if he/she suffers from type-1 or type-2 DM or from other diabetic nephropathies. 

The corresponding neighborhoods of \textit{bl\_sug} and \textit{diabetic} are highly predictive. The areas under the receiver operating characteristic (AUC-ROC) curves are $0.936$ and $0.996$ on the test data, as shown in Figure \ref{pred_uacr_egfr}c and d, respectively.

\paragraph{Gout}

\begin{figure}
\centering
\includegraphics[width=.6\textwidth]{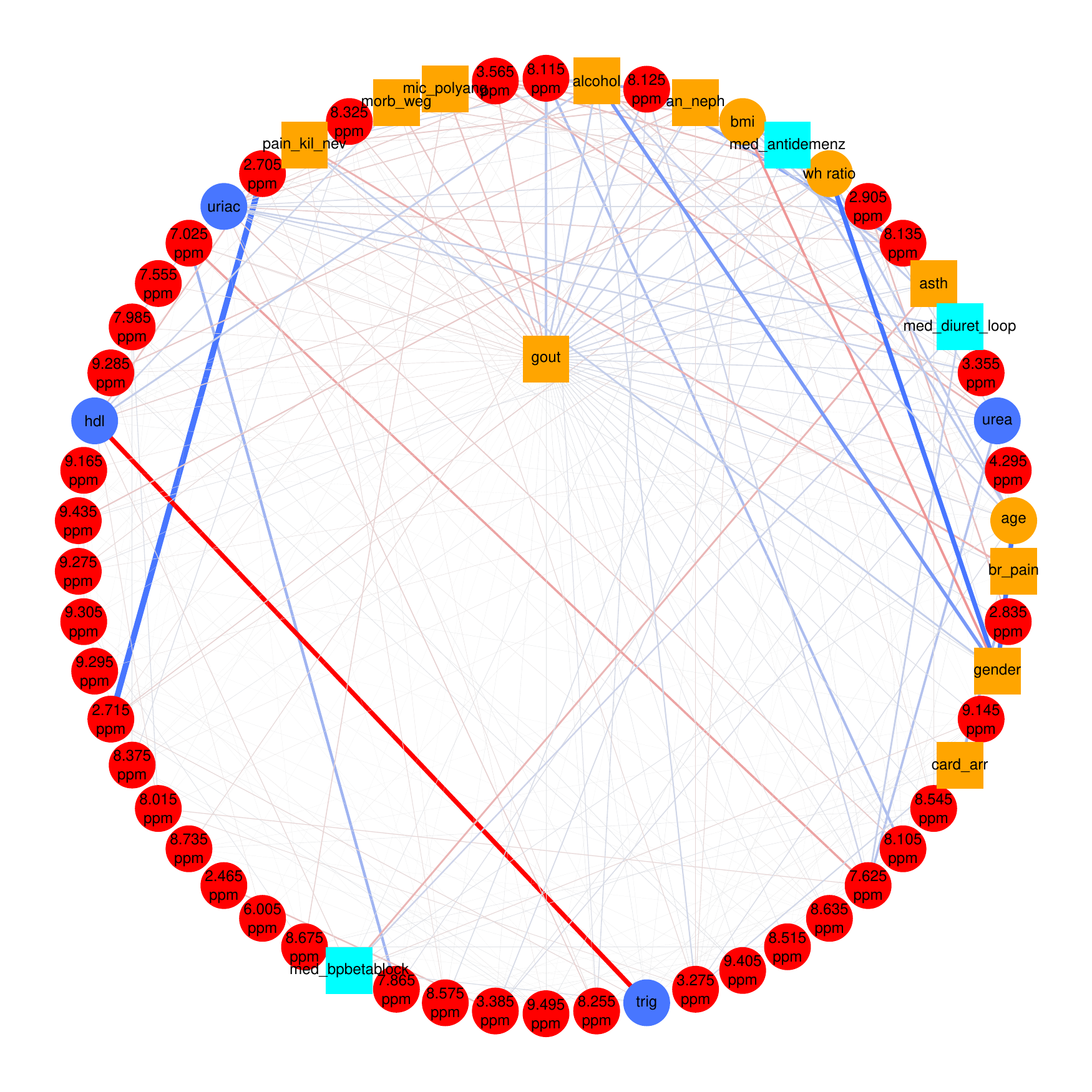}
\caption{\label{neitgout} {\footnotesize{First order neighborhood of gout ($gout$). Strong positive associations between this phenotype and the NMR bucket at 8.115 ppm (edge weight = 0.272), corresponding to unidentified small peptides, alcohol (edge weight = 0.202), 8.125 ppm (edge weight = 0.194) (unidentified small peptides), as well as analgetic nephropathy (\textit{an\_neph}) (edge weight = 0.169), and strong negative associations with the NMR bucket at 3.565 ppm (edge weight = -0.16), identified as glycine, can be observed. Gout is also connected to bmi (edge weight = 0.15), anti-dementia medication (\textit{med\_antidemenz}) (edge weight = 0.14), waist-hip ratio ($wh \; ratio$) (edge weight = 0.13), and Morbus Wegener (\textit{morb\_weg}) (edge weight = -0.13).}}}
\end{figure}

Gout is a common comorbidity in CKD patients \cite{vargas2017management, jing2014prevalence}. The first order neighborhood of gout ($gout$) as shown in Figure \ref{neitgout} comprises a number of clinical and metabolic variables. This phenotype exhibits strong positive associations with the NMR buckets at 8.115 ppm and 8.125 ppm corresponding to unidentified small peptides. An NMR bucket at 3.565ppm, which could be assigned to glycine, is strongly negatively associated with gout.

Glycine has been reported to negatively correlate with gout in other metabolic studies \cite{mahbub2017alteration}. It is involved in purine metabolism and is a precursor of uric acid \cite{mahbub2017alteration}. Decreased plasma levels of glycine in patients with gout might be caused by the increased production of uric acid in these patients \cite{mahbub2017alteration}. Note that uric acid has also been positively associated with gout in our study.

Interesting associations between gout and clinical variables are present for high alcohol consumption (positive association), analgetic nephropathy ($an\_neph$) (positive association), bmi (positive association), microscopic polyangiitis ($mic\_polyang$) (negative association), anti-dementia medication ($med\_antidemenz$) (positive association), waist-hip ratio ($wh \; ratio$) (positive association), and Morbus Wegener (negative association). Alcohol consumption, age (here positively associated with gout), male gender (here weakly positively associated with gout), loop diuretica (here positively associated), and obesity are known risk factors of gout \cite{singh2011risk, saag2006epidemiology}.  

\begin{figure*}[t!]
\centering
\includegraphics[width=.24\textwidth]{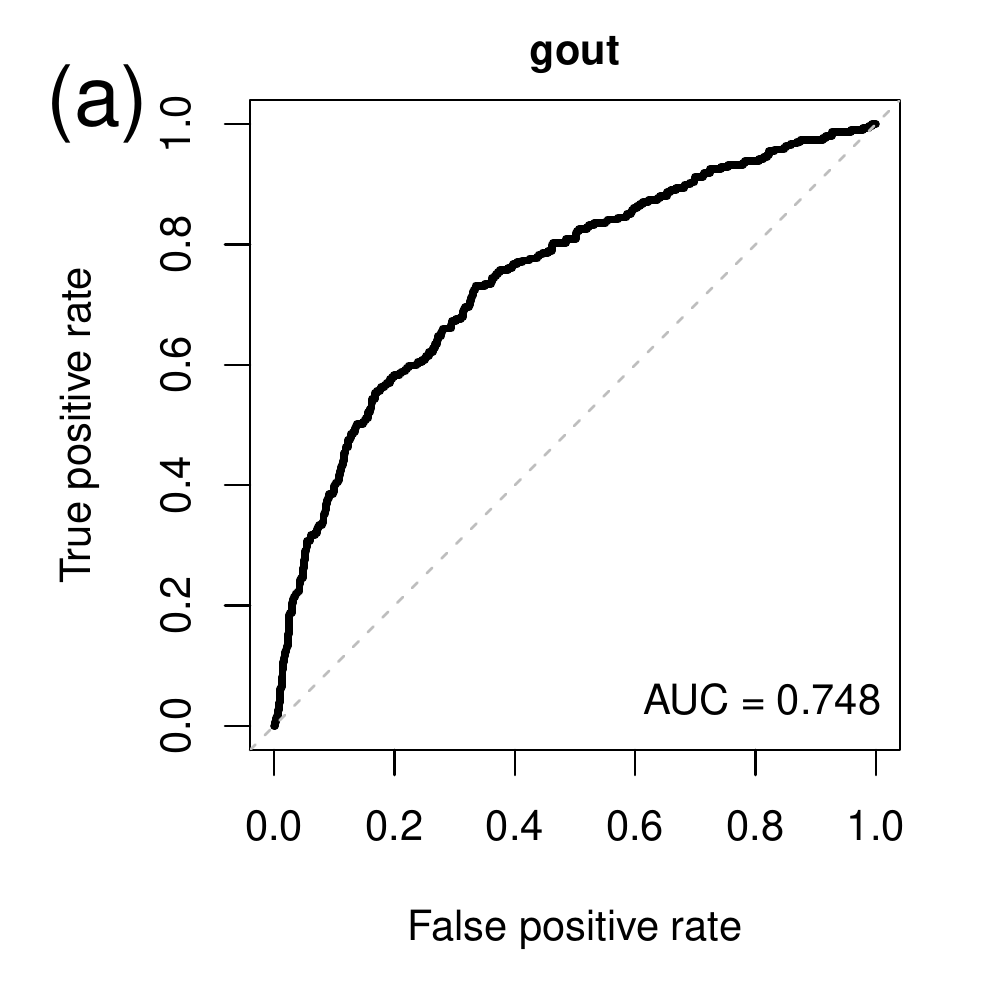}
\includegraphics[width=.24\textwidth]{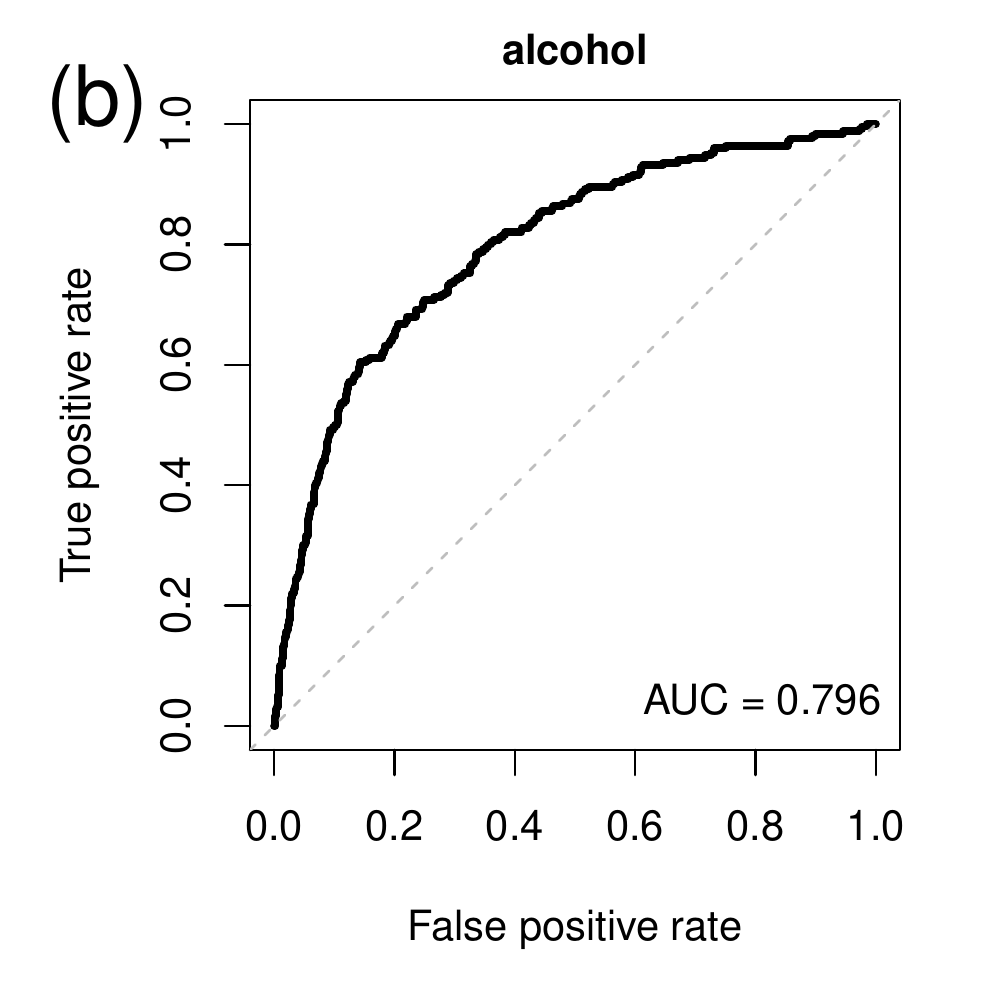}
\includegraphics[width=.24\textwidth]{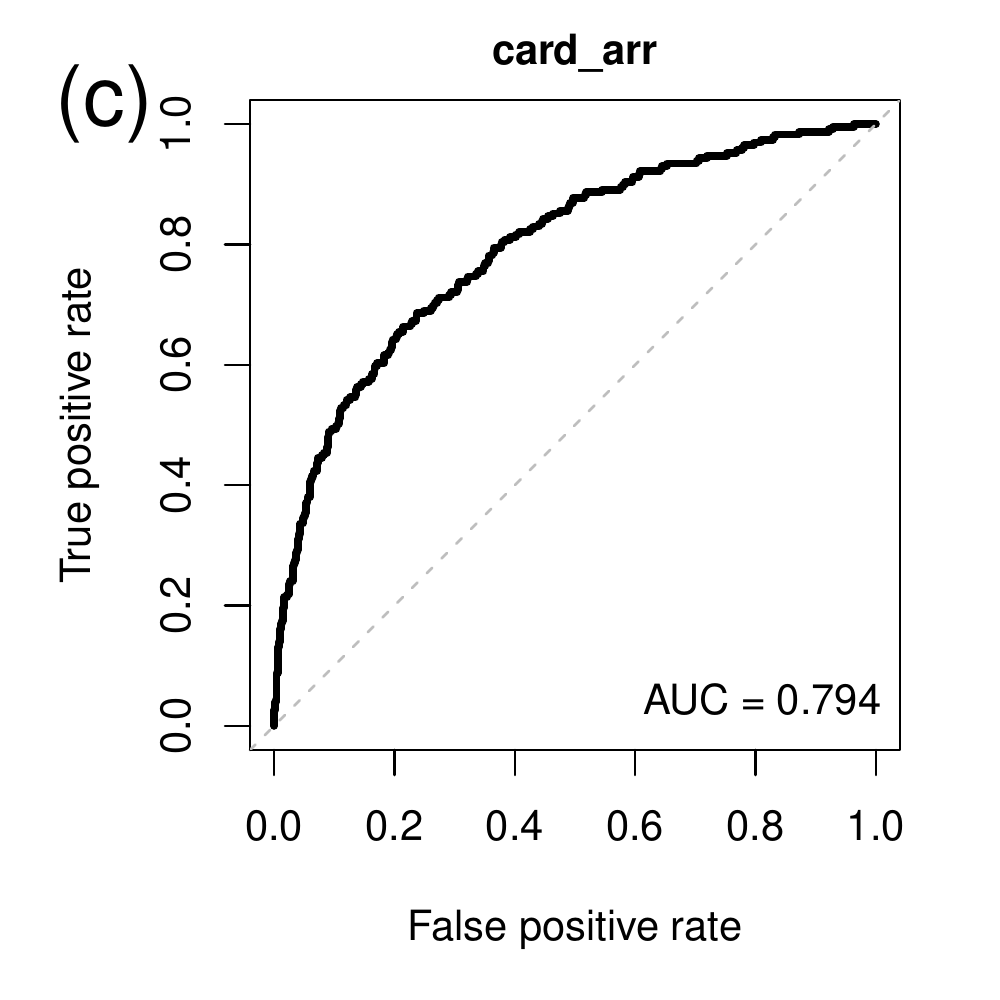}
\includegraphics[width=.24\textwidth]{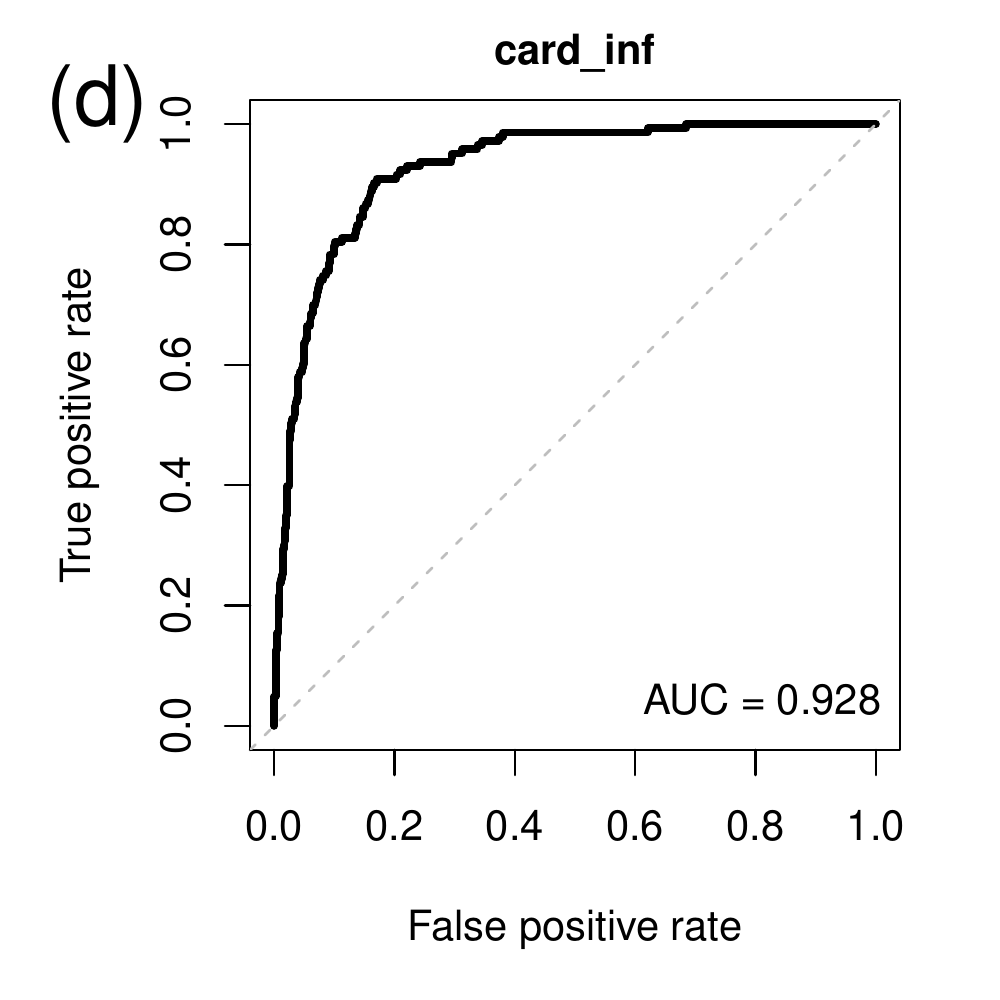}
\caption{\label{pred_gout_card}{\footnotesize{Receiver operating characteristics (ROC) curves for predicting (a) gout, (b) high alcohol consumption, (c) cardiac arrhythmia ($card\_arr$), and (d) cardiac infarction ($card\_inf$) based on their respective first order neighborhoods on independent test data. The corresponding area under the ROC curve (AUC) values are given in the lower right corners.}}}
\end{figure*}

The first order neighborhood of gout is predictive on independent test data with an AUC of 0.748, compare to Figure \ref{pred_gout_card}a.

\begin{figure}[b!]
\centering
\includegraphics[width=.6\textwidth]{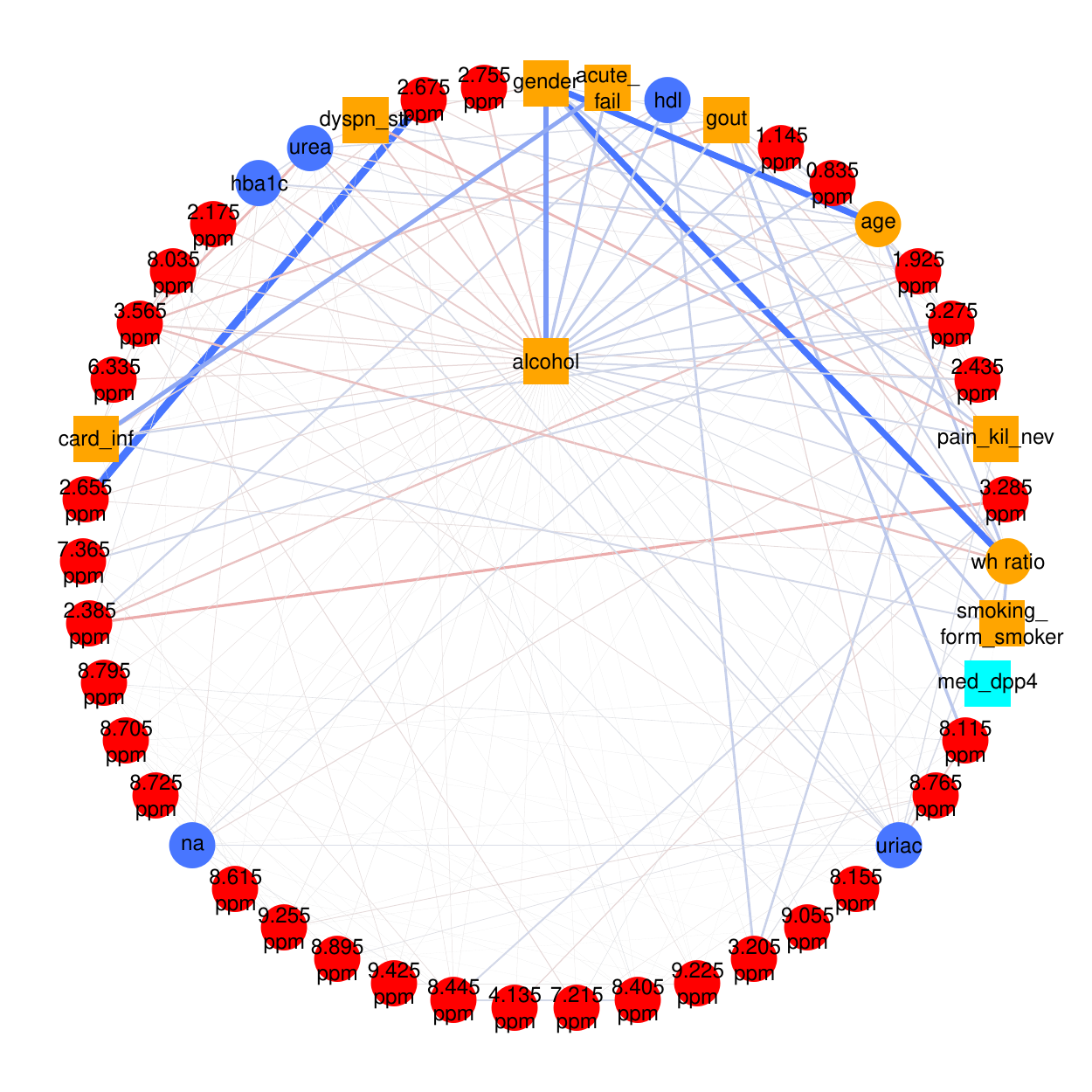}
\caption{\label{neitalc} {\footnotesize{First order neighborhood of high alcohol consumption (\textit{alcohol}). Strong associations are observed between high alcohol consumption and male gender (edge weight = 0.682), acute kidney failure ($acute\_fail$) (edge weight = 0.267), high density lipoprotein ($hdl$) values (edge weight = 0.211), gout (edge weight = 0.202), an NMR bucket at 1.145 ppm (edge weight = 0.19), containing broad lipid signals mainly identified as low density lipoprotein (LDL), an unidentified NMR bucket at 0.835 ppm (edge weight = 0.184), age (edge weight = 0.179), an unidentified NMR bucket at 2.755 ppm (edge weight = -0.138), an NMR bucket at 2.675 ppm (edge weight = -0.134), identified as citric acid, an NMR bucket at 1.925 ppm (edge weight = 0.131), identified as acetate (major signal) and minor contributions from arginine and lysine, and an NMR bucket at 3.275 ppm (edge weight = 0.127), identified as trimethylamine-N-oxide (TMAO) and minor contributions from D-glucose and betaine.}}}
\end{figure}

\paragraph{Alcohol consumption}
The assessment of alcohol consumption in the GCKD cohort was based on a questionnaire about weekly alcohol intake. Study participants, who had stated to consume alcohol not more than twice a week were classified as ``little or normal alcohol consumption'', here treated as the baseline level, and all other study participants were classified as ``high alcohol consumption''. Supplementary Figure \ref{neitalc} displays the first order neighborhood of ``high alcohol consumption''.  Interestingly, high alcohol consumption is strongly positively associated with male gender, acute renal failure ($acute\_fail$), high density lipoprotein ($hdl$), an NMR bucket at 1.145 ppm, containing broad lipid signals mainly identified as low density lipoprotein (LDL), two unidentified NMR buckets at 0.835 and 2.755 ppm, and an NMR bucket at 2.675 ppm, identified as citric acid (negatively correlated). Further, it is positively associated with an NMR bucket at 1.925 ppm, identified as acetate (major signal) and minor contributions from arginine and lysine, and an NMR bucket at 3.275 ppm, identified as trimethylamine-N-oxide (TMAO) and minor contributions from D-glucose and betaine.

Levels of large and medium-sized HDL as well as large LDL particles have been reported to be elevated by alcohol consumption \cite{e2000alcohol, mukamal2007alcohol}. It is also reported that alcohol consumption is generally higher in males than in females \cite{mukamal2007alcohol}. Citric acid has been reported to negatively correlate with alcohol consumption in other studies \cite{wurtz2016metabolic}.

As shown in Figure \ref{pred_gout_card}b, the first order neighborhood of high alcohol consumption is predictive on independent test data with an AUC of 0.796. 

\subsection{Associations in univariate screening approaches can be biased by incorrect or incomplete expert knowledge, while MGMs intrinsically account for confounding variables}

Here, we compare two different approaches to screen for associations, i.e. a standard univariate regression and the MGM data integration approach. We will illustrate that the associations discovered by the MGM are more robust for \textit{post hoc} confounder adjustment than univariately screened associations.

\subsubsection{Univariate screening}
We first screened for univariate associations between variables by regressing each variable on every single remaining variable in the dataset. We performed linear or logistic univariate regression analysis, depending on whether the response variable followed a Gaussian or a binomial distribution, respectively. For each response variable, we ranked the univariate predictor variables according to their association strength in terms of significance, here $-\log_{10}(\textrm{$p$-values})$. The predictor with the largest $-\log_{10}(\textrm{$p$-value})$ was considered as the ``top association''. Figure \ref{boxplots}a column ``top assoc.'' shows the distribution of $-\log_{10}(\textrm{$p$-values})$ of all top associations. 

Next, we repeated the regression analysis, but now in a multivariate fashion, simulating a regression analysis with confounder adjustment. We regressed each variable on its "top associated" variable and on the next five most significant predictors simultaneously, and recorded the $-\log_{10}(\textrm{$p$-value})$ of the "top associated" variable. The corresponding distribution is shown in Figure \ref{boxplots}b ``top assoc.''. We observe that the associations appear substantially weaker after variable adjustment. This trend is also visible from Figure \ref{smoothScatter}a, where we contrast adjusted (\textit{y}-axis) with unadjusted (\textit{x}-axis) $-\log_{10}(\textrm{$p$-values})$. Here, we obtained almost exclusively values in the upper half plane, see also Figure \ref{smoothScatter2}a. Finally, we regressed each response variable on its "top associated" predictor and on five additional variables randomly drawn out of the next top 10 univariate predictors. The corresponding distribution can be found in Figure \ref{boxplots}c ``top assoc.''. Again, $-\log_{10}(\textrm{$p$-values})$ of the top associations in this multivariate regression scenario appear much weaker than in the univariate screening of Figure \ref{boxplots}a.

\subsubsection{Associations in MGMs are intrinsically corrected for confounding variables}
\begin{figure}[!t]
\centering
\includegraphics[width=.8\textwidth]{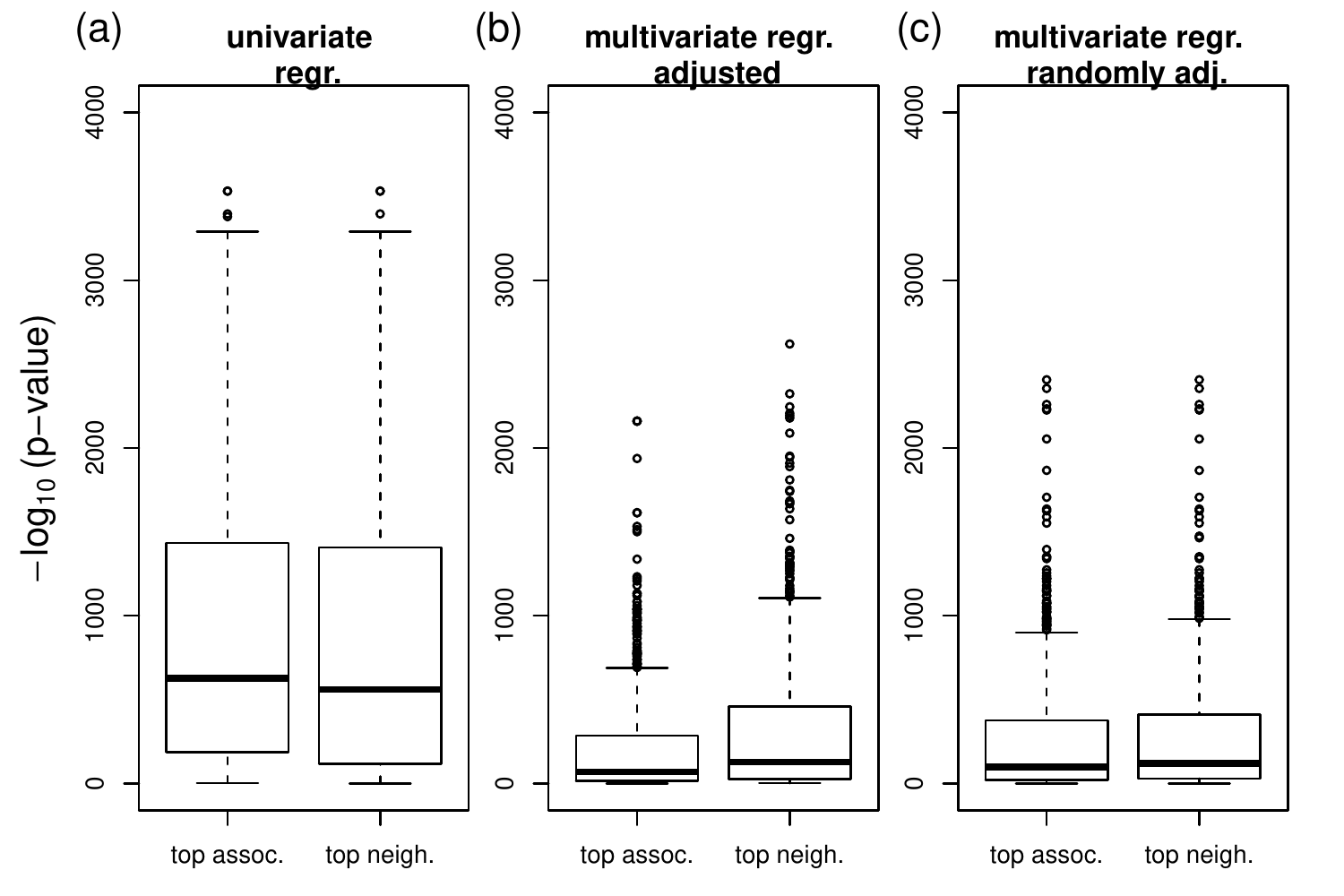}
\caption{\label{boxplots}{\footnotesize{
Effects of variable adjustment on univariate or MGM association analysis in the training set. (a) ``top assoc.'' shows the distribution of $-\log_{10}(\textrm{$p$-values})$ derived from a univariate regression. Here, we calculated $p$-values between all possible pairs of variables and collected all top associations. ``top neigh.'' shows the analogous distribution, where the top feature was selected by largest absolute edge weight in the MGM neighborhood. (b) The corresponding plot, where the $p$-values were corrected by the top five confounder variables of the univariate and MGM screening, respectively. (c) The corresponding plots, where we adjusted for the same five randomly selected features for both methods.}}}
\end{figure}

\begin{figure}[!t]
\centering
\includegraphics[width=.8\textwidth]{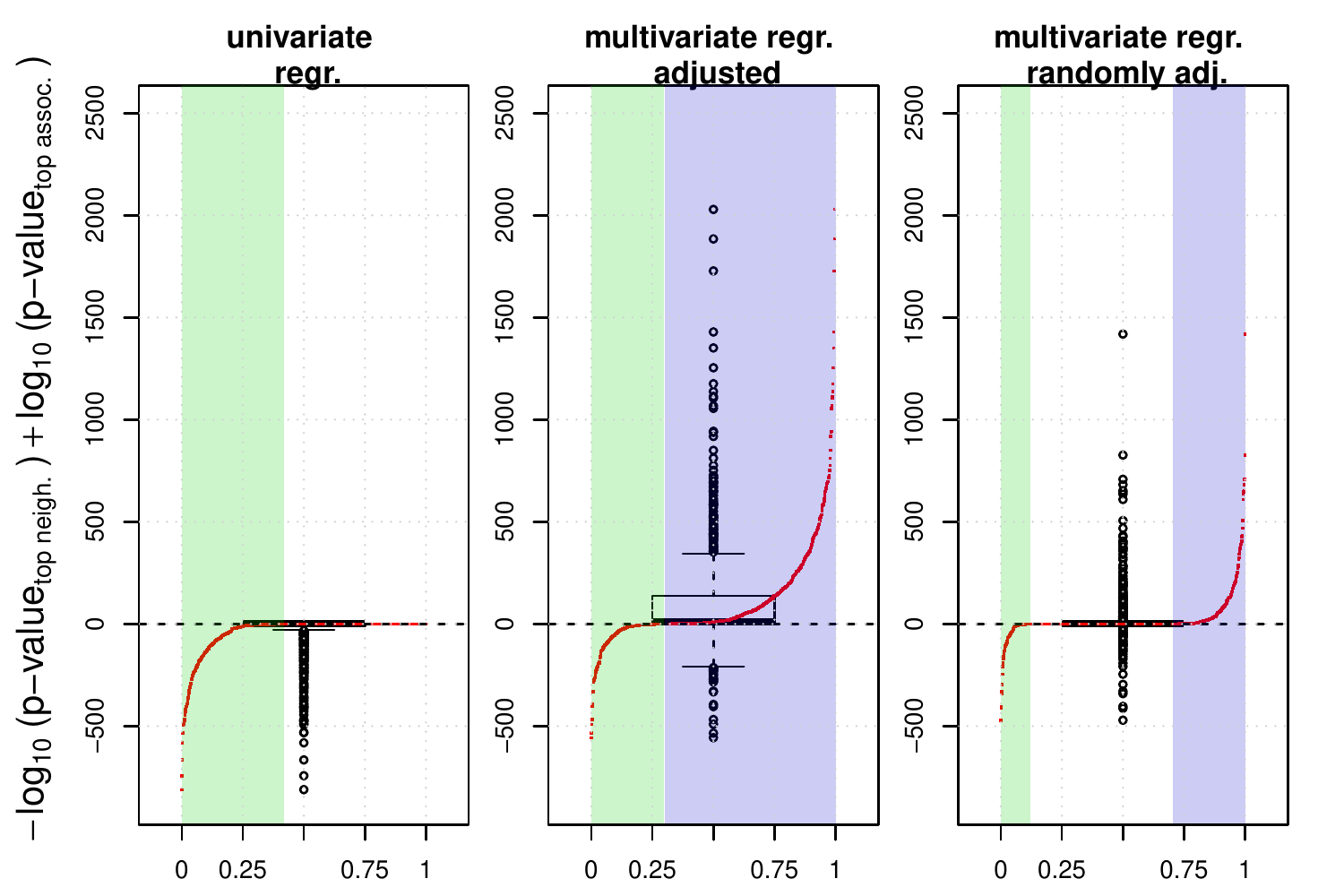}
\caption{\label{boxplots_differences}{\footnotesize{(a) to (c) show the differences between ``top neigh.'' and ``top assoc.'' in Figure \ref{boxplots}a to c, respectively: (a) shows the $-\log_{10}(\textrm{$p$-values})$ of the MGM approach minus those of the univariate screening in Figure \ref{boxplots}a, (b) shows the corresponding plot after adjusting for the respective top confounders, as shown in Figure \ref{boxplots}b, and (c) shows the corresponding plot after adjusting for the randomly selected confounders, as shown in Figure \ref{boxplots}c. The red points in each figure contrast the values on the \textit{y}-axis with their respective rank. On the \textit{x}-axis, the highest positive difference corresponds to $1$ and the most negative to $0$. The green shaded areas correspond to rank percentiles of negative, the violet shaded areas correspond to rank percentiles of positive differences, respectively.}}}
\end{figure}

We performed an analogous analysis for our MGM approach, but now, we ranked the predictors of each response variable according to their edge weights (from largest to lowest absolute value). The predictor with the largest absolute coefficient was considered as the ``top neighbor''. In a \textit{post-hoc} analysis, we regressed each response variable on its top MGM neighbor, and show the distribution of corresponding $-\log_{10}(\textrm{$p$-values})$ in Figure \ref{boxplots}a, column ``top neighbor''. Figure \ref{boxplots}b, column ``top neighbor'' shows $-\log_{10}(\textrm{$p$-values})$ for these top MGM neighbors, but now adjusted for the next top 5 neighbors in the respective MGM neighborhood. We observe that the significance decreases, as previously for the univariate screening, albeit, this effect is less pronounced (Figure \ref{boxplots}b). If we adjust for the same randomly drawn features as for the univariate approach, we obtain the results in Figure \ref{boxplots}c, which show the same trend. For Figure \ref{boxplots_differences}a to c, we subtracted the left values (``top assoc.'') from the right values (``top neighbor'') in Figure \ref{boxplots}a to c, respectively. In addition, we contrast those differences with their respective rank as red points. The $x$-axis now corresponds to the rank percentiles, starting with the most negative difference at $0$, and the highest positive difference at $1$, respectively.

In Figure \ref{boxplots_differences}a, all differences are at or below zero. Approx. 42$\%$ of the difference values are negative (see green highlighted area), and the remaining differences are zero. After confounder adjustment, Figure \ref{boxplots_differences}b, the differences are predominantly positive. Considering the rank distribution, 70$\%$ of the difference values are positive (highlighted by the violet area), and only 30$\%$ are negative (highlighted by the green area). This indicates, on average, a higher significance of the top features selected by the MGM after variable adjustment. Figure, \ref{boxplots_differences}c shows the corresponding results after an adjustment for the same randomly selected features for ``top assoc.'' and ``top neighbor''. Now, 30$\%$ of the differences are positive, only 12$\%$ are negative, and most of them are equal to zero.

\subsubsection{MGMs can reveal associations which remain hidden or underestimated in univariate approaches}
\begin{figure}[t!]
\centering
\includegraphics[width=.8\textwidth]{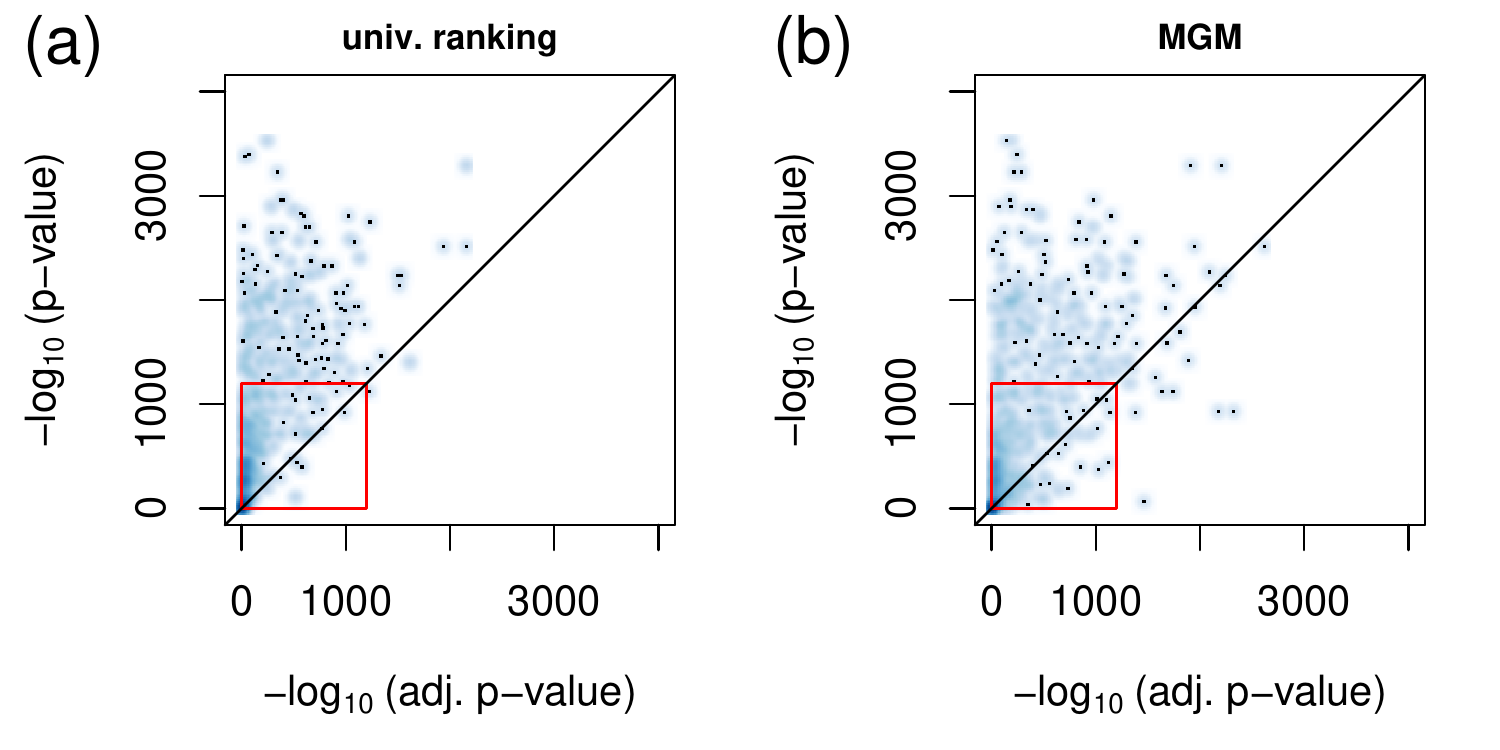}
\caption{\label{smoothScatter}\footnotesize{Smooth scatter plot of adjusted $p$-values (\textit{x}-axis) versus unadjusted (univariate) $p$-values (\textit{y}-axis) for the univariate screening (a), and the MGM (b) in the training set. The red rectangle marks an excerpt shown in detail in Figure \ref{smoothScatter2}.}}
\end{figure}

\begin{figure}[t!]
\centering
\includegraphics[width=.8\textwidth]{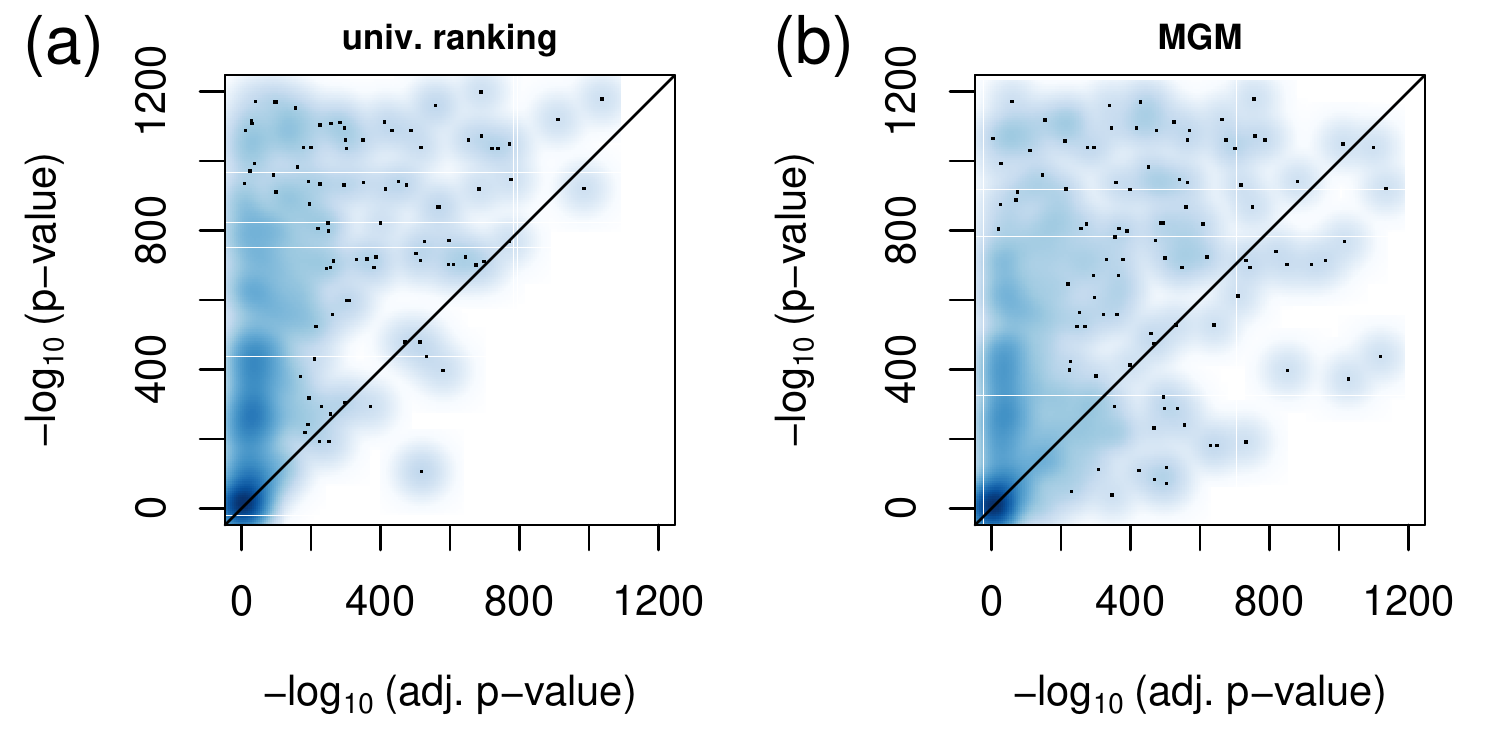}
\caption{\label{smoothScatter2}\footnotesize{The region marked by a red rectangle in Figure \ref{smoothScatter} for the univariate screening (a), and for screening by the MGM (b) in the training set. Adjusted $p$-values are shown on the \textit{x}-axis, unadjusted  $p$-values on the \textit{y}-axis.}}
\end{figure}

In Figure \ref{smoothScatter}b we contrast adjusted with unadjusted values of $-\log_{10}(\textrm{$p$-values})$ for the MGM approach. We observe that several variables are stronger associated with each other after variable adjustment compared to the unadjusted scenario. This is in contrast to the univariate approach, Figure \ref{smoothScatter}a, where almost all values were located in the upper half plane. A comparison between unadjusted and randomly adjusted $-\log_{10}(\textrm{$p$-values})$ for the univariate and the MGM approach can be found in Supplementary Figure \ref{smoothScatter_ran}.

An interesting example, how an association is revealed in a multivariate context, can be observed in the context of \textit{gout}. One of the most important associations in the neighborhood of \textit{gout} is \textit{alcohol}, which was only ranked at position $216$ in the univariate screening. Gender, in contrast, whose association is on rank $6$ in a univariate screening, appears only at position $26$ in the MGM. Thus, we can speculate that the prevalence between gout and male gender can be explained to a large extend by the higher alcohol consumption in males.

\begin{figure}[t!]
\centering
\includegraphics[width=.7\textwidth]{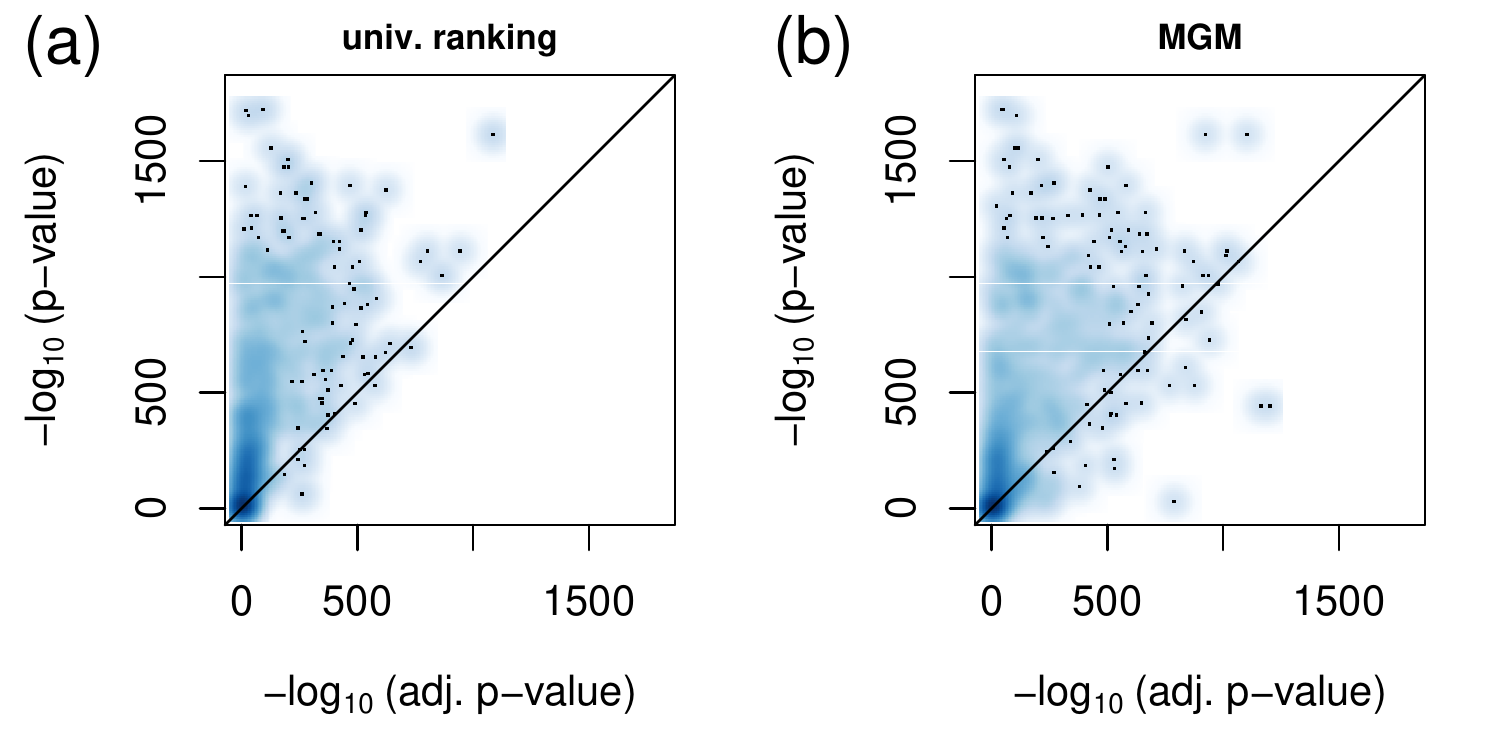}
\caption{\label{smoothScatter_te}\footnotesize{Smooth scatter plot of adjusted $p$-values (\textit{x}-axis) versus unadjusted (univariate) $p$-values (\textit{y}-axis) for the univariate screening (a), and the MGM (b) in the test set.}}
\end{figure}

Plots corresponding to Figures \ref{boxplots}, \ref{boxplots_differences}, and \ref{smoothScatter} for the validation data are shown in Supplementary Figures \ref{boxplots_te}, \ref{boxplots_differences_te}, and \ref{smoothScatter_te}, and strongly support our observations.

\subsection{TMAO is associated with cardiac infarction and cardiac arrhythmia}

\begin{figure}[h!]
\centering
\includegraphics[width=0.4\textwidth]{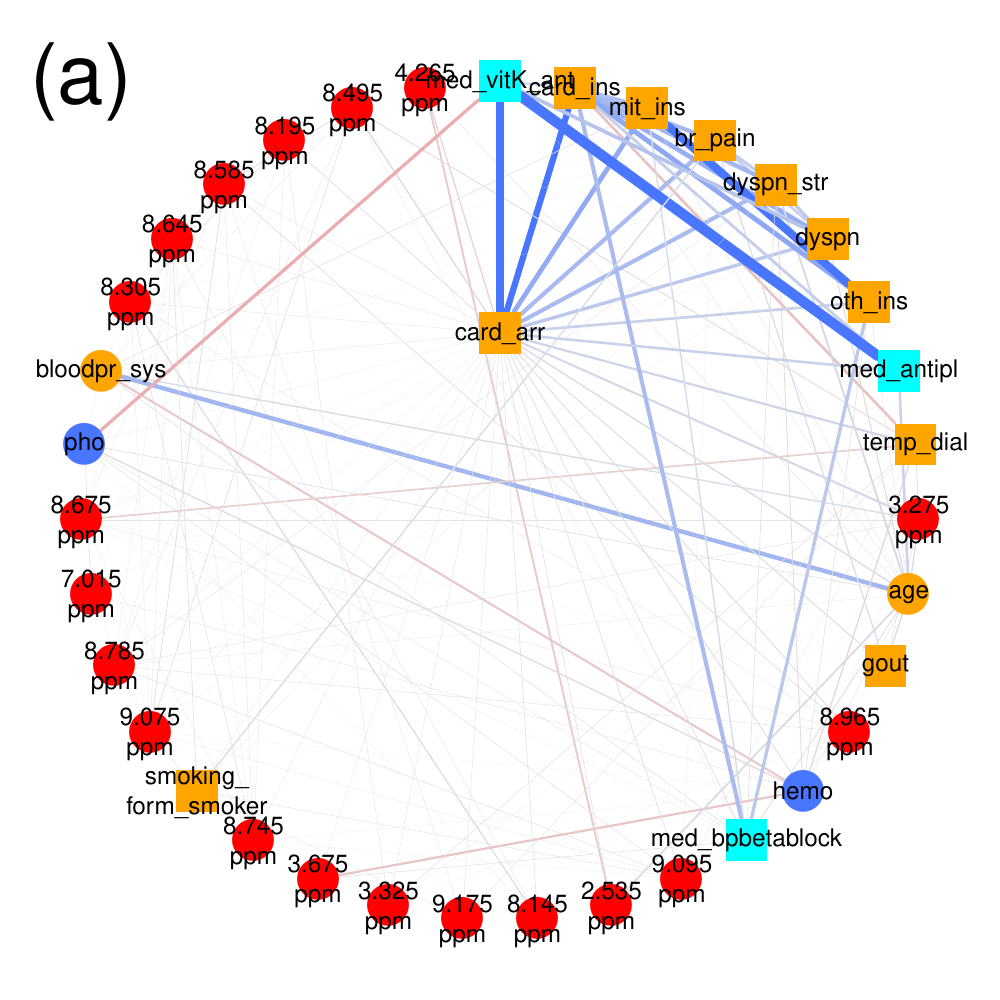}
\includegraphics[width=0.4\textwidth]{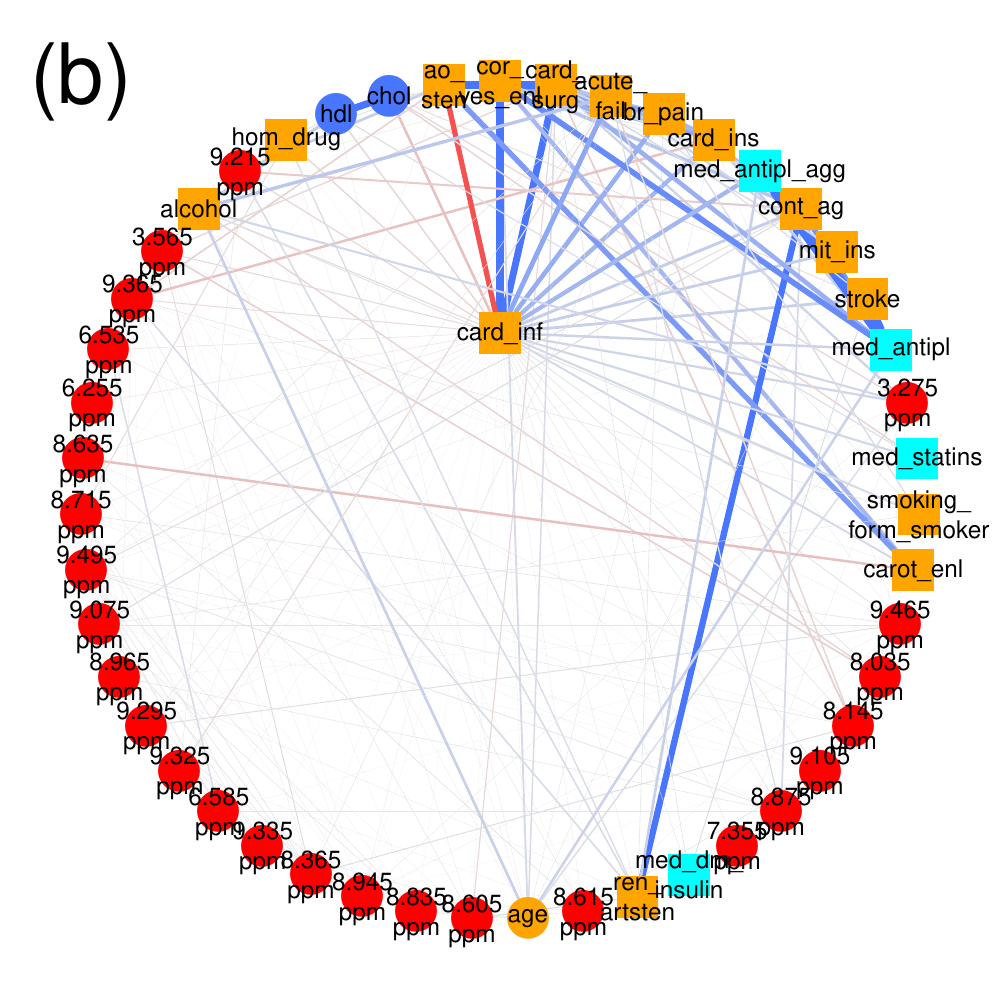}
\caption{\label{neitcard} \footnotesize{First order neighborhood of (a) cardiac arrhythmia ($card\_arr$) and (b) cardiac infarction ($card\_inf$). $card\_arr$ is strongly connected to vitamin K antagonists ($med\_vitK\_ant$) (edge weight = 1.50), heart failure (\textit{card\_ins}) (edge weight = 1.0), mitral valve insufficiency (\textit{mit\_ins}) (edge weight = 0.52), angina pectoris (\textit{br\_pain}) (edge weight = 0.39), dyspnea during physical strain ($dyspn\_str$) (edge weight = 0.38) and during the night ($dyspn$) (edge weight = 0.26), other heart valve anomalies ($oth\_ins$) (edge weight = 0.18), anti thrombotic drugs ($med\_antipl$) (edge weight = 0.17), underwent temporary dialysis (\textit{temp\_dial}), and were positively associated with an NMR bucket at 3.275ppm (edge weight = 0.12), identified as trimethylamine-N-oxide (TMAO) and minor signals of D-glucose and betaine. $card\_inf$ is strongly connected to coronary angiopathy ($cor\_ves\_enl$) (edge weight = 1.80), underwent cardiac surgery ($card\_surg$) (edge weight = 1.32), aortic valve stenosis ($ao\_sten$) (edge weight = -0.65), acute renal failure (\textit{acute\_fail}) (edge weight = 0.55), angina pectoris ($br\_pain$) (edge weight = 0.49), heart failure ($card\_ins$) (edge weight = 0.43), antiplatelet therapy ($med\_antipl\_agg$) (edge weight = 0.41), underwent catheter angiography of peripheral arteries including angioplasty of a peripheral artery ($cont\_ag$) (edge weight = 0.21), mitral valve insufficiency ($mit\_ins$) (edge weight = 0.20),  stroke ($stroke$) (edge weight = 0.19), serum cholesterol levels ($chol$) (edge weight = -0.16), anti thrombotic drugs (\textit{med\_antipl}) (edge weight = 0.16), and an NMR bucket at 3.275ppm (edge weight = 0.14). }}
\end{figure}

Cardiovascular diseases and  complications  are  typical  comorbidities  and  outcomes  in  patients with  CKD.  Here, we focus on the  first  order  neighborhood  of two  common  phenotypes in CKD patients, i.e., cardiac arrhythmia (\textit{card\_arr}) and cardiac infarction  (\textit{card\_inf}), shown in Figure \ref{neitcard}a and b, respectively.  Both variables are explained by a number of demographic and drug information parameters.  

Patients who were diagnosed with cardiac arrhythmia, compare to Figure \ref{neitcard}a, received vitamin K antagonists ($med\_vitK\_ant$) (positively associated), had experienced a heart failure (\textit{card\_ins}) (positively associated) and mitral valve insufficiency (\textit{mit\_ins}) (positively associated), experienced angina pectoris (\textit{br\_pain}) (positively associated), and dyspnea during physical strain ($dyspn\_str$) and during the night ($dyspn$) (both positively associated), got a diagnosis of other heart valve anomalies ($oth\_ins$) (positively associated), and received antithrombotic drugs ($med\_antipl$) (positively associated). 

Patients, who suffered a cardiac infarction, see Figure \ref{neitcard}b, were likely (positively associated) to have received a diagnosis of coronary angiopathy (\textit{cor\_ves\_enl}), to have undergone cardiac surgery ($card\_surg$), were less likely to have  been diagnosed with aortic valve stenosis ($ao\_sten$) (negatively associated), suffered from CKD due to acute renal failure events (\textit{acute\_fail}) (positively associated), were likely to have experienced angina pectoris ($br\_pain$) and heart failure (\textit{card\_ins}), and to have received antiplatelet therapy ($med\_antipl\_agg$). They were also likely to have underwent a catheter angiography of peripheral arteries including an angioplasty of a peripheral  artery ($cont\_ag$), to suffer from mitral valve insufficiency ($mit\_ins$), or a stroke ($stroke$), and they also had lower serum cholesterol levels ($chol$) (negatively associated).

Most interestingly, both cardiac infarction and cardiac arrhythmia were positively associated with an NMR bucket at 3.275 ppm, which could be identified as trimethylamine-N-oxide (TMAO) and minor signals from D-glucose and betaine. 

Increased plasma levels of TMAO have just recently been described as a marker for atrial fibrillation independent of hypertension, BMI, smoking, diabetes, or intake of total choline \cite{svingen2018increased}, and have been associated with a higher incidence of major adverse cardiovascular events (death, myocardial infarction, or stroke) \cite{tang2013intestinal}.

\begin{figure}[t!]
\centering
\includegraphics[width=0.5\textwidth]{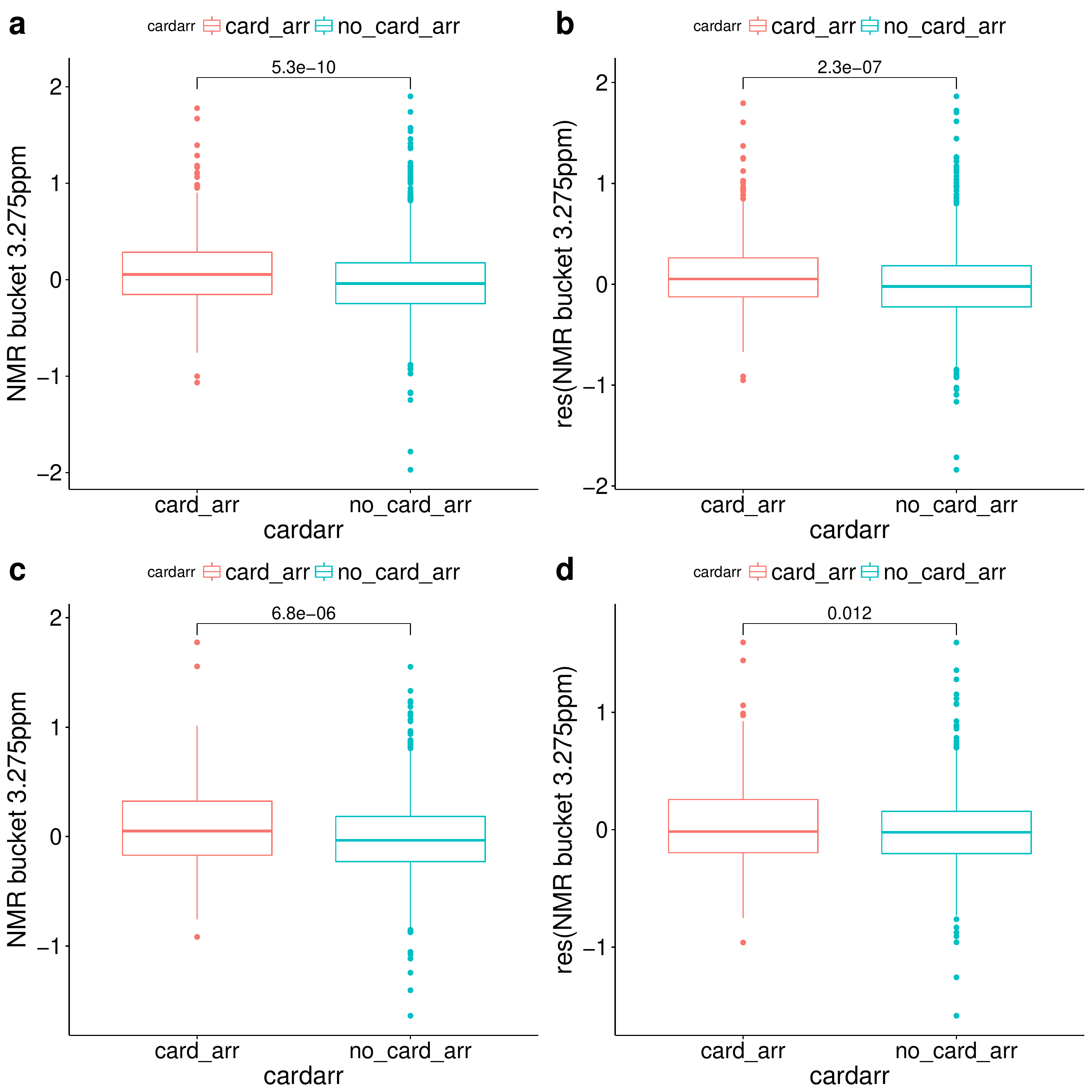}
\caption{\label{neitmaobox} \footnotesize{Boxplots comparing NMR signal intensities at 3.275ppm, which comprises NMR signals from trimethylamine-N-oxide (TMAO) (major), as well as minor signals from D-glucose and betaine. Log$_2$-transformed NMR signal intensities for patients with ($card\_arr$) and without cardiac arrhythmia ($no\_card\_arr$) for both training (a) and test data (c), respectively, are shown. The residuals of log$_2$-transformed NMR signal intensities after adjustment for BMI, systolic blood pressure, smoking status, and diabetes status for both training (b) and test data (d) are shown. $P$-values of corresponding $t$-tests are displayed above the boxes.}}
\end{figure}

Supplementary Figure \ref{neitmaobox} displays the distribution of NMR signal intensities at 3.275 ppm for both patients without and with cardiac arrhythmia for training and test data, respectively. Comparing both distributions by a Student's $t$-test yields a $p$-value of 5.3$ \cdot 10^{-10}$ and 6.8$ \cdot 10^{-6}$ for the training and test cohort, respectively. To explore possible confounding effects, we adjusted the NMR bucket intensities of 3.275 ppm for hypertension, BMI, smoking, and diabetes, which have been reported as confounders in \cite{svingen2018increased}. Intake of total choline was not accessed in the GCKD study and could therefore not be used for confounder adjustment. 
The corresponding boxplots for the residuals of NMR bucket intensities at 3.275 ppm are shown in Supplementary Figure \ref{neitmaobox}(b) and (d), respectively. The $p$-values comparing the cardiac arrhythmia with the non-cardiac arrhythmia group are still significant with values of 2.3$ \cdot 10^{-7}$ and 0.012 for training and test set, respectively.

The first order neighborhoods of cardiac arrhythmia and cardiac infarction are highly predictive on independent test data with AUCs of 0.80 and 0.93, respectively, compare to Figure \ref{pred_gout_card} (c) and (d).

\section{Discussion}

We have shown that MGMs are a versatile tool for the integration of both continuous  (Gaussian) and categorical variables.  Here, we used data collected from patients that suffered from CKD and included  variables  comprising  clinical and demographic  parameters,  medications, and blood plasma metabolites  assessed by NMR spectroscopy.  We could reveal several known relationships, such as the definitions of UACR,  eGFR,  and diabetes.   More interestingly, we observed complex  associations  between  plasma  metabolites and comorbidities like gout  and  cardiac  disorders.   Moreover, we could validate  those  associations  on test  data.

Emerging datasets provide more and more layers of information, which makes data analysis increasingly  complex.  This high complexity leaves the researcher  with an overwhelming amount of possible hypotheses,  which require  extensive  tests  and  validations.  Thus,  there  is an urgent need for automated tools  that hint   towards  interesting observations.  Here,  MGMs  are predestined, because they  condense the available  data  into graphs that researchers  can easily read and interpret. Moreover,  MGMs consider the  whole system  of variables  and measurements simultaneously.  As a consequence, they correct for confounding factors, such as age, gender, and medications, automatically.  Thus,  if the  MGM observes an interesting association,  it is likely not  a bystander effect of other  variables  if those  are part  of the  analysis  itself.  We showed this in a scenario where we corrected for artificially selected confounders. Particularly, we could demonstrate that confounder adjustment can be necessary to reveal associations, as we exemplified for the association of gout with high alcohol consumption. Finally, we illustrated that the MGM discovered an association  of TMAO  with cardiac  arrhythmia, which remains significant after adjustment for variables selected by expert knowledge.

MGMs are undirected graphical models.  Thus,  they do not include information  about  causality. In the context of Gaussian  graphical models, there are plenty of works that estimate  causal relations from observational data  only \cite{kalisch2007estimating, maathuis2009estimating,maathuis2010predicting}. Combining those methods  with MGMs could further strengthen our data  integration approach.

In summary,  we tested  MGMs for the  integrative analysis  of categorical  and  continuous variables in the  context  of CKD.  We illustrated its application, provided  software  to  estimate   MGMs,  and  exemplified  their  interpretation.  Moreover, we have reported interesting associations  between  plasma metabolites and the CKD comorbidities  gout and cardiac  disorders.

\section{Acknowledgement}

This work was supported in part by the German Federal Ministry of Education and Research (BMBF grant no. 01 ER 0821). 

GCKD investigators are:
University of Erlangen-N\"urnberg, Germany: Kai-Uwe Eckardt, Stephanie Titze, Hans-Ulrich Prokosch, Barbara Bärthlein, André Reis, Arif B. Ekici, Olaf Gefeller, Karl F. Hilgers, Silvia H\"ubner, Susanne Avenda$\tilde{\textrm{n}}$o, Dinah Becker-Grosspitsch, Nina Hauck, Susanne A. Seuchter, Birgit Hausknecht, Marion Rittmeier, Anke Weigel, Andreas Beck, Thomas Ganslandt, Sabine Knispel, Thomas Dressel, Martina Malzer\\
Technical University of Aachen, Germany: J\"urgen Floege, Frank Eitner, Georg Schlieper, Katharina Findeisen, Elfriede Arweiler, Sabine Ernst, Mario Unger, Stefan Lipski\\
Charité, Humboldt-University of Berlin, Germany: Elke Schaeffner, Seema Baid-Agrawal, Kerstin Petzold, Ralf Schindler\\
University of Freiburg, Germany: Anna K\"ottgen, Ulla Schultheiss, Simone Meder, Erna Mitsch, Ursula Reinhard, Gerd Walz\\
Hannover Medical School, Germany: Hermann Haller, Johan Lorenzen, Jan T. Kielstein, Petra Otto\\
University of Heidelberg, Germany: Claudia Sommerer, Claudia F\"ollinger, Martin Zeier\\
University of Jena, Germany: Gunter Wolf, Martin Busch, Katharina Paul, Lisett Dittrich\\
Ludwig-Maximilians University of M\"unchen, Germany: Thomas Sitter, Robert Hilge, Claudia Blank\\
University of W\"urzburg, Germany: Christoph Wanner, Vera Krane, Daniel Schmiedeke, Sebastian Toncar, Daniela Cavitt, Karina Sch\"onowsky, Antje B\"orner-Klein\\
Medical University of Innsbruck, Austria: Florian Kronenberg, Julia Raschenberger, Barbara Kollerits, Lukas Forer, Sebastian Sch\"onherr, Hansi Wei\ss ensteiner\\
University of Regensburg, Germany: Peter J. Oefner, Wolfram Gronwald, Helena Zacharias\\
Department of Medical Biometry, Informatics and Epidemiology (IMBIE), University of Bonn: Matthias Schmid

\newpage

\section{Supplementary materials and methods}

\subsection{Cohort description}

The study cohort comprises 3705 participants of the German Chronic Kidney Disease (GCKD) study, whose detailed baseline clinical and demographic characteristics are described elsewhere \cite{eckardt2011german, titze2014disease}. Written declarations of consent had been obtained from all study participants before inclusion. From each patient, one EDTA-plasma specimen had been collected at the baseline time-point and stored at -80$^\circ$C until NMR measurement.

\subsection{Clinical and demographic variables}

For this study, we included 17 clinical chemistry parameters measured by SYNLAB International GmbH (Munich, Germany) and Central Lab, University Hospital Erlangen, Germany \cite{titze2014disease}, 73 demographic parameters including age, sex, disease history and lifestyle factors, and 46 different drug treatments, resulting in a total of 136 clinical variables. Supplementary Table S1 provides a list of all included  variables, their corresponding distributions, as well as assessment information.

\subsection{NMR spectroscopy}

For NMR measurements, 400 $\mu$L of unfiltered EDTA-plasma were mixed with 200 $\mu$L of 0.1 mol/L phosphate buffer at pH 7.4, 50 $\mu$L of 0.75\% (w) 3-trimethylsilyl-2,2,3,3-tetradeuteropropionate (TSP) dissolved in deuterium oxide, and 10 $\mu$L of 81.97 mmol/L formic acid (all from Sigma-Aldrich, Taufkirchen, Germany), the latter serving as internal standard for referencing and quantification \cite{wallmeier2017quantification}.
NMR experiments were carried out on a 600 MHz Bruker Avance III (Bruker BioSpin GmbH, Rheinstetten, Germany) employing a triple-resonance ($^1$H, $^{13}$C $^{31}$P, $^2$H lock) cryogenic probe equipped with z-gradients and an automatic cooled sample changer. Since the presence of broad protein signals might hinder the interpretation of metabolite signals, 1D $^1$H spectra employing a Carr-Purcell-Meiboom-Gill (CPMG) pulse-sequence for the suppression of macromolecular signals were acquired for each EDTA-plasma specimen. For each 1D $^1$H CPMG spectrum, 128 scans were collected into 73728 data points employing presaturation during relaxation for water suppression. Four dummy scans were acquired prior to measurement, the spectral width was 20.02 ppm, the relaxation delay was 4 s and the acquisition time amounted to 3.07 s. The filtering delay amounted to 0.08 s, resulting in a total acquisition time of about 16 min per sample.\\
NMR signals were assigned to known metabolites  by comparison with reference spectra of pure compounds acquired under equal experimental conditions employing the Bruker Biofluid Reference Compound Database BBIOREFCODE 2-0-3 \cite{u2013current}.

\subsection{Data preprocessing}
\label{dataPrepro}
All continuous clinical and demographic variables except for age, systolic and diastolic blood pressure were log$_2$ transformed to remove heteroscedasticity. Recoding of discrete variables is detailed in Supplementary Table S1.  In summary, we included 25 continuous and 111 discrete clinical and demographic variables, respectively.\\
All NMR spectra were referenced with respect to the formic acid signal at 8.463 ppm. Since signal positions between spectra may be subject to minor shifts due to slight differences in pH, salt concentration, and/or temperature between samples, equidistant binning was employed to compensate for these effects. The spectral region from 9.5 - 0.5 ppm was evenly split into 900 bins of 0.01 ppm width using Amix VIEWER 3.9.13 (Bruker BioSpin GmbH, Rheinstetten, Germany). All spectral data were scaled relative to the formic acid signal from 8.5 - 8.45 ppm in order to reduce variations in spectrometer performance. NMR buckets corresponding to the remainders of the suppressed water signal as well as minor contributions from urea (6.0 - 4.5 ppm), formic acid (8.47 - 8.45 ppm), as well as dimethylamine (2.73 - 2.72 ppm) have been excluded prior to data analysis, resulting in a total number of 743 NMR buckets.\\
For further analysis, data were imported into $R$ version 3.2.1 (Development Core Team 2009). To minimize heteroscedasticity of the NMR data, bucket intensities were log$_2$ transformed and additionally subjected to mean-value subtraction.\\
All continuous variables were scaled to standard units.  

\subsection{Mixed Graphical Models}
\label{MGMs}
\subsubsection{Probability density function}
Integration of all discrete and continuous variables is achieved by estimating the conditional dependencies between them by a Mixed Graphical Model (MGM) \cite{lauritzen1996graphical}. MGMs are undirected probabilistic graphical models, where each node corresponds to one variable, and the edges between two nodes represent a conditional dependency between them given all other variables in the graphical model. If there exists no edge between two nodes, these two variables are conditionally independent of each other given all other variables in the MGM.

Lee and Hastie \cite{pmlr-v31-lee13a} proposed the following probability density function $p \left( x,y;\Theta \right)$ to describe a pairwise graphical model on continuous and discrete variables:
\begin{align}
\label{probdenfunc}
p \left( x,y;\Theta \right) &\propto \exp \Big(-\frac{1}{2} \sum_{s=1}^p\sum_{t=1}^p \beta_{st} x_{s} x_{t}+\sum_{s=1}^{p}\alpha_{s} x_{s} \nonumber\\&+\sum_{s=1}^p \sum_{j=1}^{q} \rho_{sj}(y_{j})x_{s}+\sum_{j=1}^q \sum_{r=1}^q \phi_{rj}(y_r , y_j)\Big).
\end{align}
Here, $x_s$ denotes the $s$th of $p$ continuous variables, $y_j$ denotes the $j$th of $q$ discrete variables with $L_j$ states, $\beta_{st}$ represents the continuous - continuous edge, and $\alpha_s$ the continuous node potential, respectively, $\rho_{sj}$ is the continuous - discrete edge potential, represented as a vector of size $L_j$, $\phi_{rj}$ is the discrete - discrete edge potential, and $\Theta = \left[ \{ \beta_{st} \}, \{ \alpha_s \}, \{ \rho_{sj} \}, \{ \phi_{rj} \} \right]$ summarizes the whole parameter space. $-\frac{1}{2} \sum_{s=1}^p\sum_{t=1}^p \beta_{st} x_{s} x_{t}+\sum_{s=1}^{p}\alpha_{s} x_{s}$ describes conditional dependencies between two continuous variables $x_s$ and $x_t$, corresponding to the probability density function of a Gaussian Graphical Model (GGM) \cite{pmlr-v31-lee13a}. If $\beta_{st} = 0$, no edge between $x_s$ and $x_t$ appears in the MGM, indicating an independency between these two nodes conditioned against all other nodes in the graphical model. Analogously, $\sum_{j=1}^q \sum_{r=1}^q \phi_{jr}(y_j , y_r)$, which corresponds to a discrete pairwise Markov Random Field \cite{pmlr-v31-lee13a}, describes conditional dependencies between two discrete variables $y_j$ and $y_r$ with $L_j$ and $L_r$ states, respectively. If all entries of the matrix $\phi_{rj}$ are equal to zero, the two discrete variables are conditionally independent given all others. Finally, $\sum_{s=1}^p \sum_{j=1}^{q} \rho_{sj}(y_{j})x_{s}$ gives the conditional dependencies between a continuous variable $x_s$ and a discrete variable $y_j$, represented by a vector of size $L_j$. If all entries of this vector are equal to zero, the two variables are conditionally independent given all other variables in the MGM. In summary, if the whole parameter space $\Theta = \left[ \{ \beta_{st} \}, \{ \alpha_s \}, \{ \rho_{sj} \}, \{ \phi_{rj} \} \right]$ is determined, we are able to fully describe all conditional dependencies between all variables in the considered system, here the population under investigation in the GCKD study, which can be represented in a network.

\subsubsection{Pseudo-log-likelihood, LASSO penalty, and algorithmic implementation}
A maximum likelihood estimate (MLE) of Eq. (\ref{probdenfunc}) requires a summation over all discrete states, which is numerically only feasible for small numbers of discrete variables. As an alternative, we use the pseudo-likelihood method of \cite{pmlr-v31-lee13a}, where the negative pseudo log-likelihood is given by
\begin{align}
\mathcal{L}(\Theta|x,y)=-\sum_{s=1}^{p} \log{p(x_s|x_{\backslash s},y;\Theta)}-\sum_{r=1}^{q} \log{p(y_{r}|x,y_{\backslash r};\Theta)}\,.
\label{eq:negpl}
\end{align} 
Here, $p(x_s | x_{\backslash s}, y;\Theta)$ is the conditional distribution of a Gaussian variable $x_s$, and $p(y_r| y_{\backslash r,}, x;\Theta)$ is the conditional distribution of a discrete variable $y_r$ with $L_r$ states:
\begin{align}
&p(x_s | x_{\backslash s}, y;\Theta)=\nonumber\\&\frac{\sqrt{\beta_{ss}}}{{\sqrt{2 \pi}}}\exp{ \left(\frac{-\beta_{ss}}{2} \left(\frac{\alpha_s + \sum_{j} \rho_{sj} (y_j) - \sum_{t \neq s} \beta_{st} x_{t} }{\beta_{ss}} -x_{s}\right)^2\right)}\nonumber\,,\\
&p(y_r| y_{\backslash r,}, x;\Theta)=\nonumber\\&\frac{\exp{\left(\sum_{s} \rho_{sr}(y_r) x_{s} +\phi_{rr} (y_r,y_r) +\sum_{j\neq r} \phi_{rj}(y_r,y_{j}) \right)}}{\sum_{l=1}^{L_r } \exp{\left(\sum_{s} \rho_{sr}(l) x_{s} +\phi_{rr} (l,l) +\sum_{j\neq r} \phi_{rj}(l,y_{j}) \right)} } \,.\label{eq:discond}
\end{align}
Note, that the pseudo-log-likelihood method does not require a summation over the whole state space, which makes it computationally more tractable than MLE.

We augmented the pseuod-log-likelihood estimate of (\ref{eq:negpl}) by a LASSO penalty to calibrate overfitting 
\begin{align}
\lambda \left(\sum_{s=2}^p\sum_{t=1}^{s} \omega_{st} |\beta_{st}| +\sum_{s=1}^p \sum_{j=1}^q \omega_{sj} |\rho_{sj}|_{\tilde{1}} +\sum_{j=2}^q \sum_{r=1}^{j} \omega_{rj}|\phi_{rj}|_{\tilde{1}} \right)\,.
\label{eq:penpl}
\end{align}
Here, we used the definitions
\begin{equation}
|\phi_{rj}|_{\tilde 1}=\sum^{L_r}_{l=1}\sum^{L_j}_{m=1}|\phi_{rj}(l,m)|\,\quad \mbox{and}\quad |\rho_{j}|_{\tilde 1}=\sum^{L_j}_{l=1}|\rho_{j}(l)|\,.
\end{equation}
The weights $w_{st}$, $w_{sj}$, and $w_{rj}$ were chosen as
\begin{align}
w_{st}=\sigma_{s} \sigma_{t}\,,\quad
w_{sj}=\sigma_{s} \sqrt{ \sum_{a} p_{b} (1-p_{b})}\,,\nonumber\\
w_{rj}=\sqrt{ \sum_{a} p_{a} (1-p_{a}) \sum_{b} q_{b} (1-q_{b})}\,,
\end{align}
where $\sigma_{s}$ is the standard deviation of the continuous variable $x_{s}$ and $p_{a} = Pr(y_{r}=a)$ and $q_b= Pr(y_j =b)$ \cite{pmlr-v31-lee13a}. For the multinomial variables, our parameter space is redundant \cite{pmlr-v31-lee13a}. This was taken into account by multiplying the penalty factors which correspond to baseline levels by a factor of 10. This forces their couplings to zero.

Optimization problem Eq. (\ref{eq:penpl}) is convex \cite{pmlr-v31-lee13a} and can be efficiently solved by proximal algorithms. We implemented our software in Python, with computationally expensive parts written in C. Here, we used the Python package \textit{apgpy} accessed under \url{https://github.com/bodono/apgpy}, which performs the proximal gradient descent with Nesterov acceleration and adaptive restarts \cite{o2015adaptive}.

\subsubsection{Calibration of the penalty parameter $\lambda$}
We chose the penalty parameter $\lambda$ according to the Extended Bayesian Information Criterion \cite{chen2008extended,foygel2010extended}
\begin{equation}
\label{ebic}
\textrm{EBIC}_{\gamma}(\textbf{E})=-2l_n(\hat{\Theta}(\textbf{E}))+|\textbf{E}|\log n + 4 |\textbf{E}|\gamma \log P,
\end{equation}
where $\gamma$ is a parameter that can be chosen in the interval $[0,1]$, $n$ is the number of measurements and $P$ is the number of variables. $\textbf{E}$ is the set of edge parameters unequal to zero, and $l_n(\hat{\Theta} (\textbf{E}))$ the corresponding maximized pseudo-log-likelihood. For calculating both $P$ and $\textbf{E}$ we removed the baseline levels from the adjacency matrix. The minimum of Eq. (\ref{ebic}) determines the penalty parameter $\lambda$, which we estimated by line search. This line search sweeps the $\lambda$-sequence $\lambda_0 \cdot 2^{2,1.75,1.5,1.25,\ldots,-4.5,-4.75,-5}$, with $\lambda_0 = \sqrt{\frac{\log(p + q)}{n}}$. Throughout the article, we discuss the model $\gamma=1$, which corresponds to the sparsest model according to EBIC. For visualization, we used the R packages \textit{qgraph} \cite{epskamp2012qgraph} and \textit{igraph} \cite{igraph2006}, where we transformed the edge weights by the formula $f(x)=\log(1+10x)$.

\subsection{Association analyses}
Analysis were performed using the statistical analysis software R version 3.3.3. For logistic regression, we used the R package \textit{logistf} \cite{logistf}.
\newpage
\section{Supplementary Figures}

\begin{figure}[b!]
\centering
\includegraphics[width=.7\textwidth]{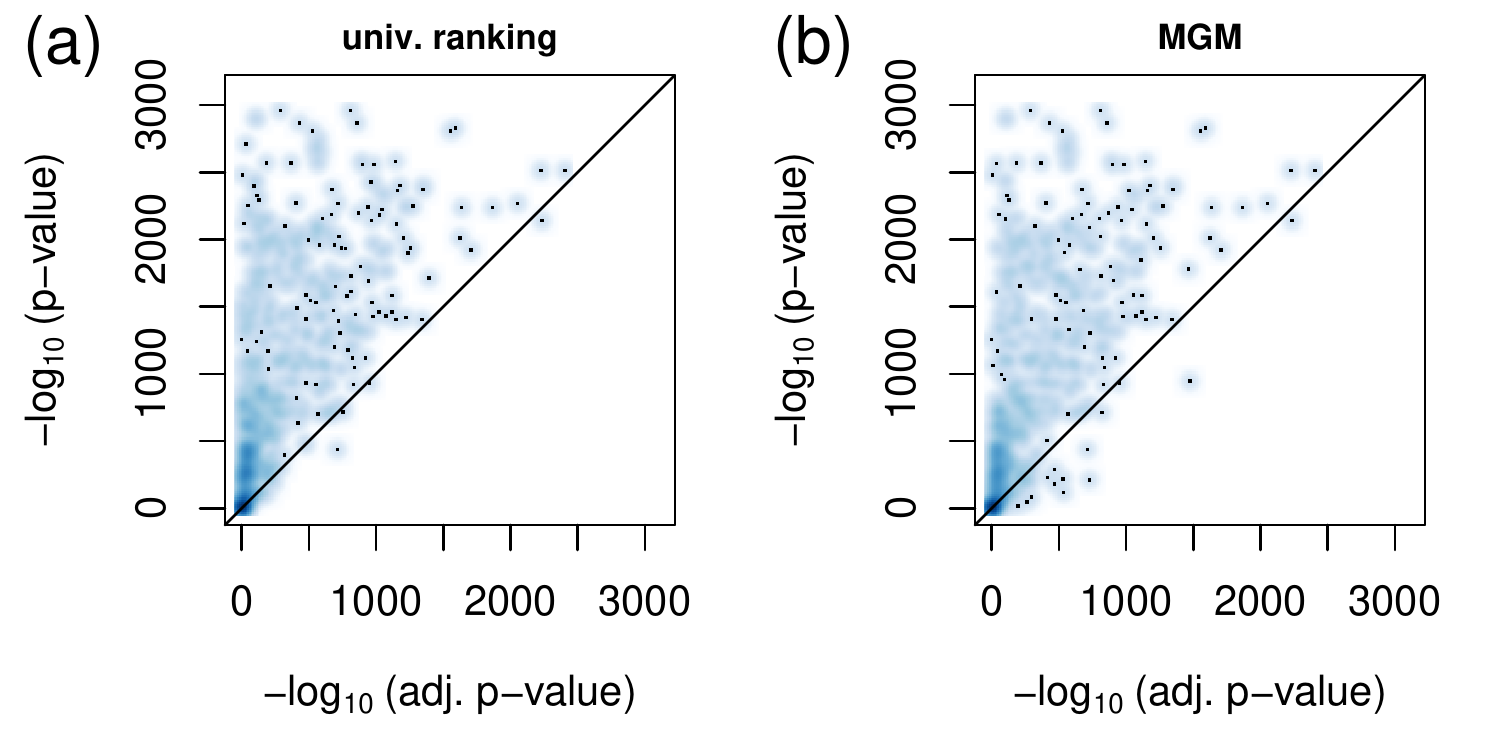}
\caption{\label{smoothScatter_ran}\footnotesize{Smooth scatter plot of randomly adjusted $p$-values (\textit{x}-axis) versus unadjusted (univariate) $p$-values (\textit{y}-axis) for the univariate screening (a), and the MGM (b) in the test set.}}
\end{figure}

\begin{figure}[!b]
\centering
\includegraphics[width=.7\textwidth]{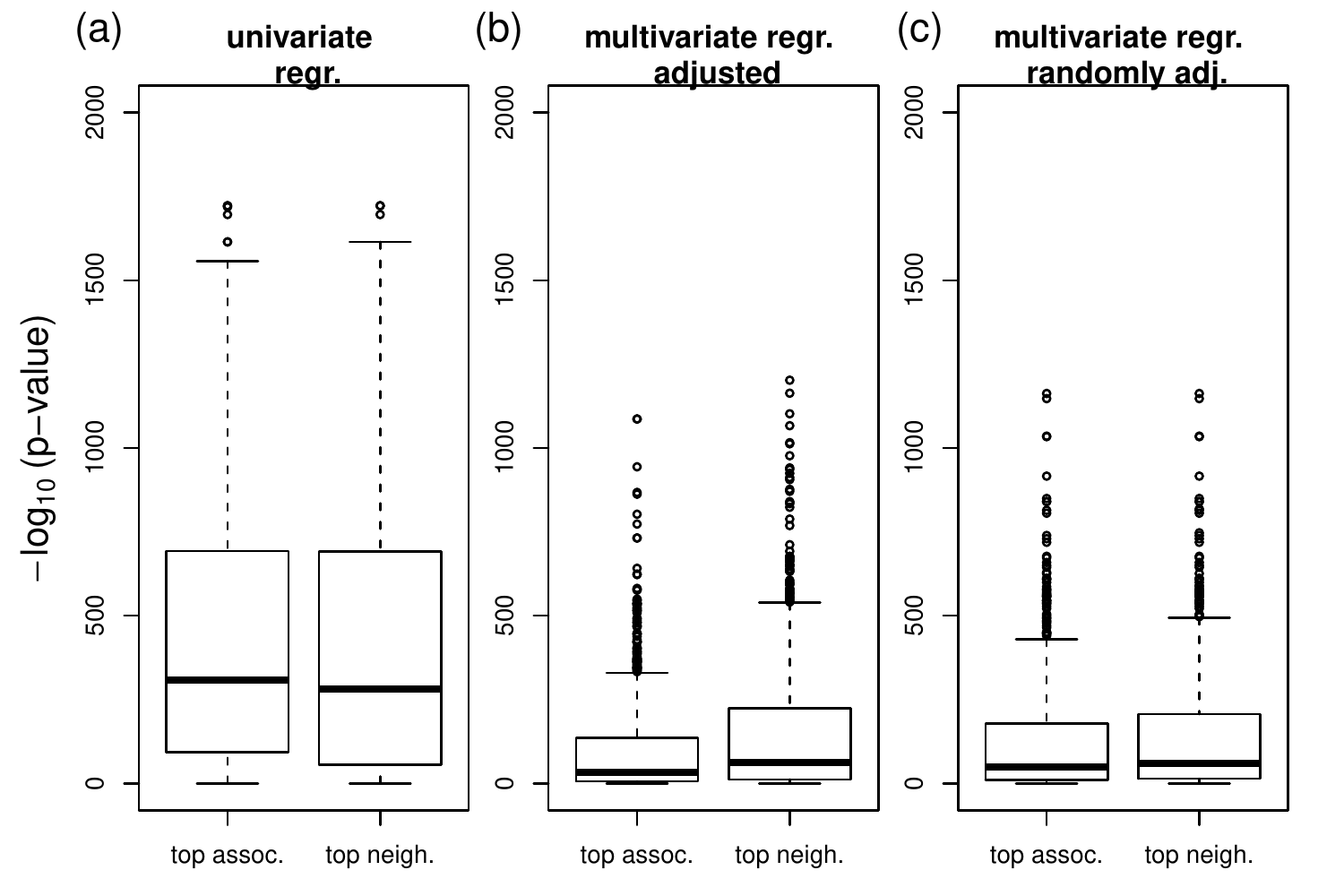}
\caption{\label{boxplots_te}\footnotesize{Effects of variable adjustment on univariate or MGM association analysis in the test set. (a) ``top assoc.'' shows the distribution of $-\log_{10}(\textrm{$p$-values})$ derived from a univariate regression. Here, we calculated $p$-values between all possible pairs of variables and collected all top associations. (a) ``top neigh.'' shows the analogous distribution, where the top feature was selected by largest absolute regression weight in the MGM neighborhood. (b) The corresponding plot, where the $p$-values were corrected by the top five confounder variables of the univariate and MGM screening, respectively. (c) The corresponding plot, where we adjusted for the same five randomly selected features for both methods.}}
\end{figure}

\begin{figure}[!t]
\centering
\includegraphics[width=.7\textwidth]{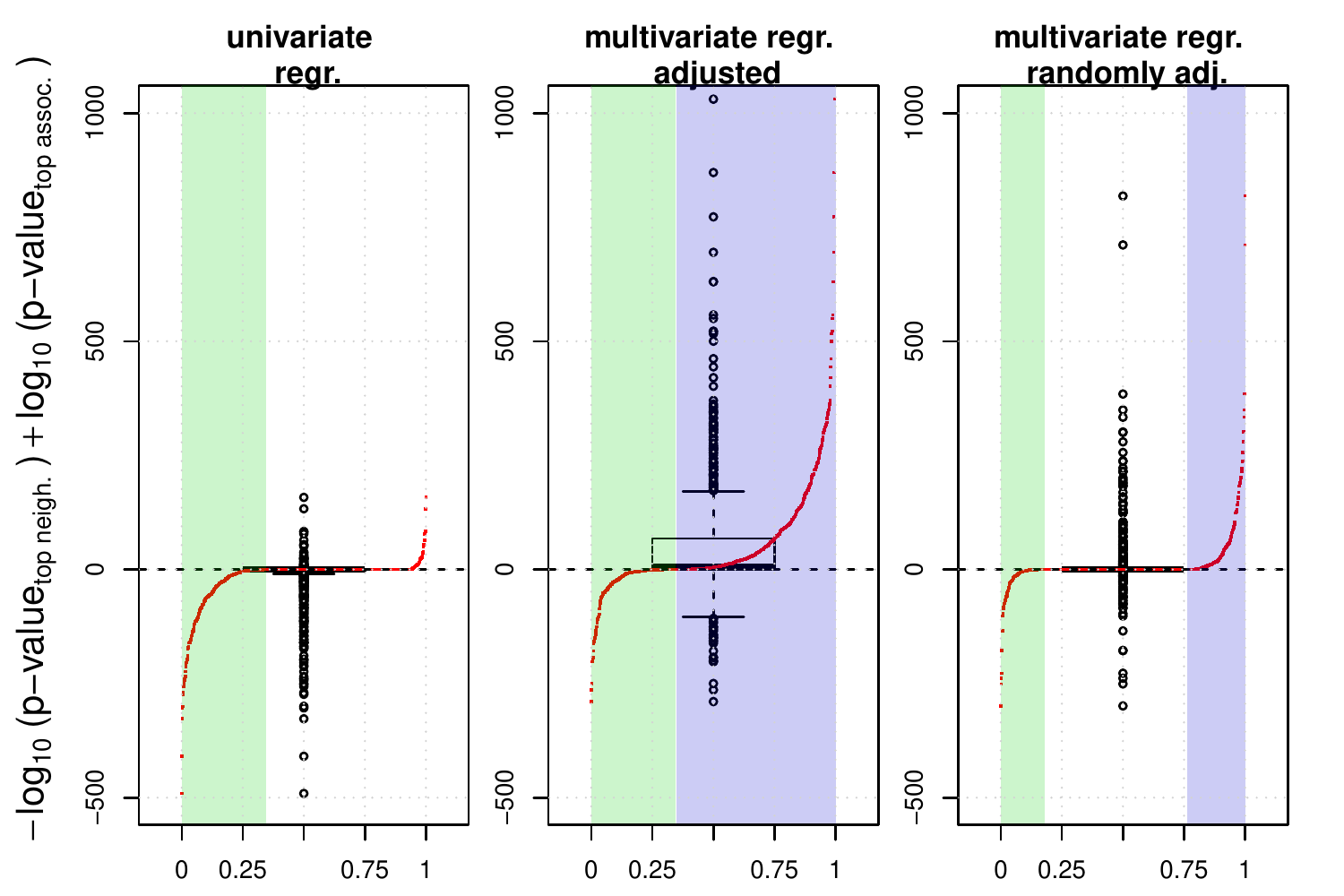}
\caption{\label{boxplots_differences_te}\footnotesize{(a) to (c) show the differences between ``top neigh.'' and ``top assoc.'' in Figure \ref{boxplots_te}a to c, respectively: (a) shows the $-\log_{10}(\textrm{$p$-values})$ of the MGM approach minus those of the univariate screening in Figure \ref{boxplots_te}a, (b) shows the corresponding plot after adjusting for the respective top confounders, as shown in Figure \ref{boxplots_te}b, and (c) shows the corresponding plot for adjusting for the randomly selected confounders, as shown in Figure \ref{boxplots_te}c. The red point in each figure contrast the values on the \textit{y}-axis with their respective rank. On the \textit{x}-axis, the highest positive difference correspond to $1$ and the most negative to $0$. The green shaded areas correspond to rank percentiles of negative, the violet shaded areas correspond to rank percentiles of positive differences.}}
\end{figure}

\end{document}